\documentclass[aps,prl,twocolumn,amssymb,superscriptaddress]{revtex4}
\usepackage{subfigure}
\usepackage{graphicx}
\usepackage{rotating} 
\usepackage{color}
\usepackage{lscape}
\usepackage{rotating}
\usepackage{amsmath,amsthm,placeins}
\usepackage{epstopdf}
\begin{document}

\title{Contact statistics highlight distinct organizing principles of
proteins and RNA}


\author{Lei Liu}
\affiliation{%
School of Computational Sciences, Korea Institute for Advanced Study, 85 Heogiro Dongdaemun-Gu, Seoul 02455, Republic of Korea} 
\author{Changbong Hyeon}
\thanks{To whom correspondence should be addressed. Email: hyeoncb@kias.re.kr}
\affiliation{%
School of Computational Sciences, Korea Institute for Advanced Study, 85 Heogiro Dongdaemun-Gu, Seoul 02455, Republic of Korea} 


\begin{abstract}
Although both RNA and proteins have densely packed native structures, chain organizations of these two biopolymers are fundamentally different.   
Motivated by the recent discoveries in chromatin folding that interphase chromosomes have territorial organization with  signatures pointing to metastability, we analyzed the biomolecular structures deposited in the Protein Data Bank and found that the intrachain contact probabilities, $P(s)$ as a function of the arc length $s$, decay in power-law $\sim s^{-\gamma}$ over the intermediate range of $s$, $10\lesssim s\lesssim 110$. 
We found that the contact probability scaling exponent is $\gamma\approx 1.11$ for large RNA ($N>110$), $\gamma\approx 1.41$ for small sized RNA ($N<110$), and $\gamma\approx 1.65$ for proteins. 
Given that Gaussian statistics is expected for a fully equilibrated chain in polymer melts, 
the deviation of $\gamma$ value from $\gamma=1.5$ for the subchains of large RNA in the native state 
suggests that the chain configuration of RNA is not fully equilibrated. 
It is visually clear that folded structures of large sized RNA ($N\gtrsim 110$) adopt crumpled structures, partitioned into modular multi-domains assembled by proximal sequences along the chain, whereas the polypeptide chain of folded proteins looks better mixed with the rest of the structure.  
Our finding of $\gamma\approx 1$ for large RNA might be an ineluctable consequence of the hierarchical ordering of the secondary to tertiary elements in the folding process.
\end{abstract}

\maketitle


RNA and proteins, under an appropriate environmental conditions, adopt three-dimensionally compact native folds that are essential for a variety of biological functions. 
Despite general similarities of the folding principles that both biopolymers are made of sequences foldable to a functionally competent structure as an outcome of evolutionary selection \cite{schuster1994PRSL,Tinoco99JMB,Thirum05Biochem,Chen00PNAS,morcos2014PNAS}, the overall shape of the native RNA differs from that of proteins in several aspects. 
Proteins are in general more compact, globular, and flexible than RNA \cite{Hyeon06JCP_2}. 
Such differences may be originated from distinct nature of building block.  
The energy scale of binary interaction that pairs nucleotides is typically greater than that of amino acids. 
Furthermore, the requirement of charge neutralization (or screening) along the backbone differentiates the foci of RNA dynamics, especially at the early stage of folding \cite{ThirumARPC01}, from those of proteins. 

Spotlighted in the recent studies of chromatin folding exploiting fluorescence in-situ hybridization (FISH) \cite{langer1982PNAS,cremer2001NRG} and chromosome conformation capture techniques \cite{Dekker2002Science,lieberman09Science,dekker2013NRG}, human chromosomes in the interphase have a territorial organization \cite{cremer2001NRG} and the individual chromosome is also partitioned into a number of topologically associated domains (TADs), possibly mediated by proteins such as CTCF and cohesin \cite{zuin2014PNAS}. 
The contact probability $P(s)$ of two loci separated by the genomic distance $s$ can provide glimpses into the arrangement of chromatin chain. 
From the polymer perspective, a test chain in a ``fully" equilibrated homogeneous polymer melt is expected to  
obey the Gaussian statistics because of the screening of excluded volume interaction \cite{GrosbergBook}, thus satisfying $P(s)\sim s^{-3/2}$. 
It was, however, shown that $P(s)$ of human chromatin in cell nucleus displays $P(s)\sim s^{-1.08}$ at the genomic scales of $1<s<10$ Mb \cite{lieberman09Science}. 
To account for the origins of the human genome organization and its characteristic scaling exponent $\gamma=1.08$ and patterns of contact map demonstrating TADs, several different models have been put forward, which include the crumpled (fractal) globule \cite{lieberman09Science,grosberg1988JP, Mirny11ChromoRes}, random loop \cite{2007Bohn051805}, strings and binders switch model \cite{Barbieri12PNAS}, and confinement-induced glassy dynamics \cite{Kang2015PRL}. 

Besides the overall shape, chain organizations of the native folds of RNA and proteins are in general visually different from each other. 
Compared with proteins in which $\alpha$-helices, $\beta$-strands, and loops thread through one another to form a native structure, a folded RNA with large $N$ looks more crumpled; a number of secondary structure elements (helices, bulges,  loops) forming \emph{independently stable} modular contact domains are further assembled into a compact three dimensional structure. 
Here, borrowing the several statistical measures that have been used to study the genome/chromosome organization inside cell nucleus, we substantiate the fundamental differences between the chain organizations of RNA and proteins in native states and discuss its significance in connection to their folding mechanisms. 
\\

\section*{METHODS}
\begin{figure*}[ht]
\begin{center}
\centerline{\includegraphics[width=1.5\columnwidth]{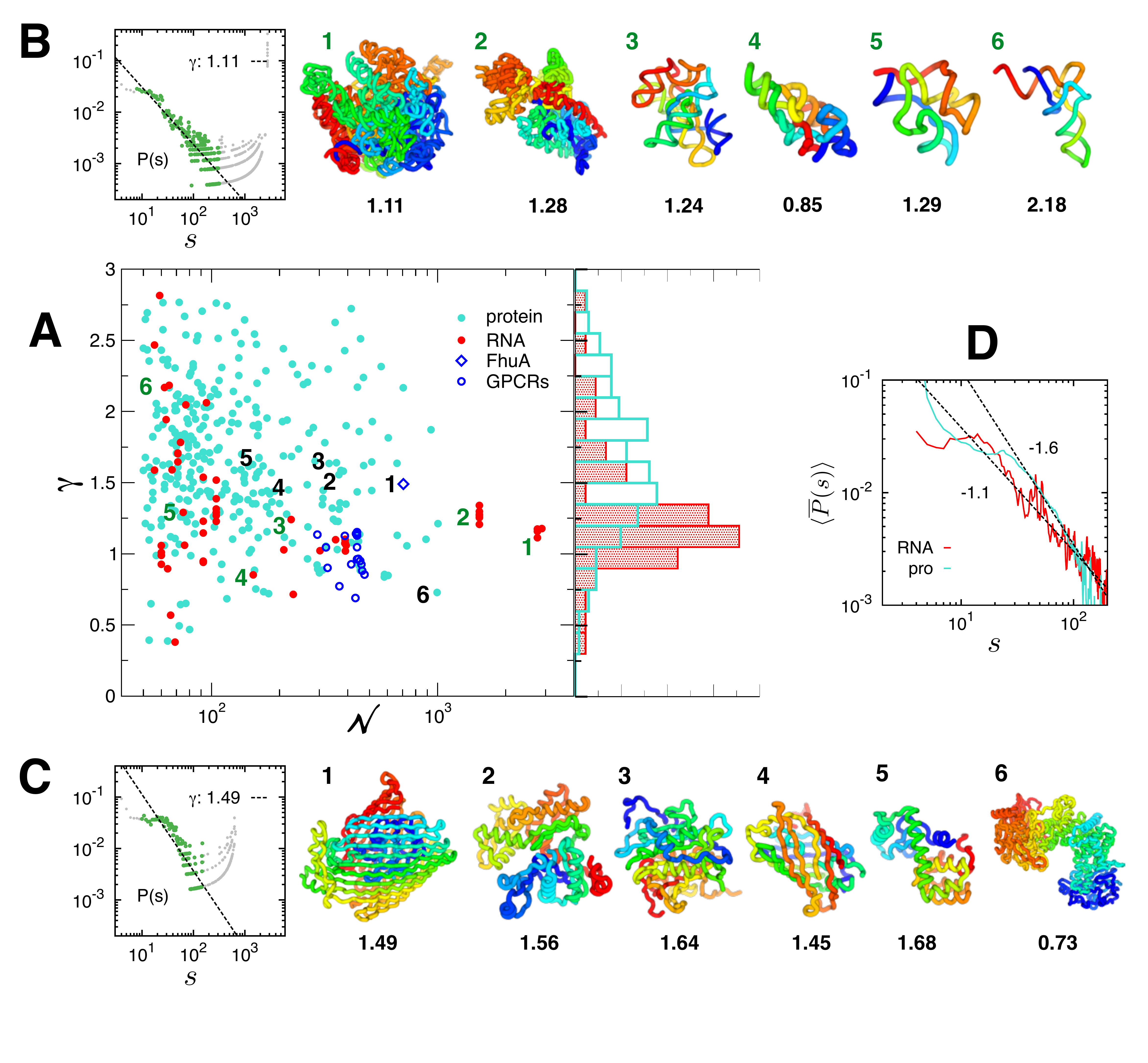}}
\caption{
Contact probability scaling exponent, $\gamma$, and chain configurations of RNA and proteins. 
{\bf A.} $\gamma$ versus $N$ obtained for 60 individual RNA (data in red) and 324 proteins (data in cyan), whose $\gamma$ value is obtained from the power-law fit to $P(s)$ with the correlation coefficient (c.c.) greater than  $>0.9$ (see Fig.S4 for $\gamma$ vs. $N$ plot with error bars, and Tables I, II in Supporting Material for PDB entries used here).  
The data points for FhuA and GPCRs are included for further discussion although c.c$<0.9$.   
Histograms of $\gamma$, $p(\gamma)$ for RNA ($\gamma\approx 1.30\pm 0.44$) and proteins ($\gamma=1.65\pm 0.50$) are shown on the right. 
{\bf B.} Representative structures of RNA in rainbow coloring scheme from 5' (blue) to 3' (red), indexed with the number in $\gamma$ vs. $N$ plot.  
Depicted are the structures of 
{\bf 1.} a large subunit of rRNA (PDB entry: 2O45 \cite{Pyetan2007PAC}, $\gamma=1.11$), 
{\bf 2.} a small subunit of rRNA  (2YKR, $\gamma=1.28$) \cite{guo2011PNAS}, 
{\bf 3.} Twort group I ribozyme (1Y0Q, $\gamma=1.24$) \cite{Golden2005NSMB}, 
{\bf 4.} A-type ribonuclease P (1U9S, $\gamma=0.85$) \cite{krasilnikov2004Science}, and 
{\bf 5.} TPP-riboswitch (3D2G, $\gamma=1.29$) \cite{Thore2008JACS}.
$P(s)$, which provides $\gamma$ value, is shown for a large subunit of rRNA on the left corner. The scaling exponent ($\gamma$) of $P(s)\sim s^{-\gamma}$ is obtained from the fit (dashed line) to the data points in green; the data in gray are excluded from the fit (see Supporting Material for details of fitting procedure). 
{\bf C.} Protein structures in rainbow coloring scheme from the N- (blue) to C-terminus (red).  
Depicted in (C) are the structures of 
{\bf 1.} FhuA (1QJQ $\gamma=1.49$) \cite{ferguson2000ProteinSci}, 
{\bf 2.} an actin monomer (1J6Z $\gamma=1.56$) \cite{2001Otterbein708}, 
{\bf 3.} metacaspase (4AF8, $\gamma=1.6$) \cite{2012McLuskey7469}, 
{\bf 4.} Green Fluorescent Protein (1EMA, $\gamma=1.4$) \cite{1996Ormo1392}, 
{\bf 6.} T4 lysozyme (2LZM, $\gamma=1.7$) \cite{1987Weaver189,shank2010Nature},
and 
{\bf 6.} Chondroitin Sulfate ABC lyase I (1HN0, $\gamma=0.73$) \cite{Huang2003JMB}. 
$P(s)$ for FhuA is shown on the left corner. 
{\bf D.} The mean contact probabilities, $\langle \overline{P}(s)\rangle$, calculated over the RNA and protein structures in the PDB.
}
\label{structures}
\end{center}
\end{figure*}

\noindent{\bf Calculation of contact probability and extraction of scaling exponent.}
Using atomic coordinates of RNA and protein from Protein Data Bank (PDB), 
we consider that two residues $i$ and $j$ are in contact if the minimum distance between any two heavy atoms of these residues, located at $\vec{r}_i$ and $\vec{r}_j$, is smaller than a cut-off distance $d_{c}$ (= 4 \AA).  
The contact probability for a biomolecule $\alpha$ with chain length $N_{\alpha}$ (the number of residues) is thus determined by calculating  
\begin{equation}
P_{\alpha}(s) = \frac{\sum_{i<j}^{N_{\alpha}}\delta(|i-j|-s) \Theta(d_{c}-\min|\vec{r}_{i}-\vec{r}_{j}|)}{\sum_{i<j}^{N_{\alpha}} \delta(|i-j|-s)},
\label{Ps}
\end{equation}
where $\Theta(x) = 1$ for $x\ge 0$; otherwise $\Theta(x) = 0$. 
Two examples of $P(s)$ are given in Figs.\ref{structures}B and \ref{structures}C. 
The power law relation of $P(s) \sim s^{-\gamma}$ is observed over the intermediate scale. 
We determined the value of $\gamma$ by fitting $P(s)$ over the range of $s_{\text{min}}<s<s_{\text{max}}$. 
The details of fitting procedure are discussed in the Supporting Material.
\\

\noindent{\bf Mean contact probability. }
Each structure in the PDB has different chain size $N_{\alpha}$ ($\alpha=1,2,\ldots,I_{\text{max}}$). 
Thus, to consider the non-uniform distribution of chain size in computing the mean contact probability, 
we calculated the following $N$-dependent probability averaged over the total number of distinct chain sizes:
\begin{align} 
\langle \overline{P}(s)\rangle = \frac{1}{N_{\text{max}}-N_{\text{min}}+1}\sum_{N=N_{\text{min}}}^{N_{\text{max}}}\overline{P}(s|N),
\label{meancontactprob}
\end{align}
where $\overline{P}(s|N)\equiv \sum_{{\alpha}=1}^{I_{\text{max}}}\delta (N_{\alpha}-N)P_{\alpha}(s)/\sum_{{\alpha}=1}^{I_{\text{max}}}\delta(N_{\alpha}-N)$ 
is the mean contact probability for the structures with chain size $N$, and we used the value of $P_{\alpha}(s)$ only for the range of $4 \leq s\leq N_{\alpha}^{2/3}$.
$\langle \overline{P}(s)\rangle$ for RNA and proteins are shown in Fig. \ref{structures}D. 
$\langle M(s)\rangle$, $\langle n_s(s)\rangle$, $\langle \text{DOP}\rangle$, $\langle\text{DOS}\rangle$, and $\langle R(s)\rangle$ were calculated using similar definitions as Eq.~\ref{meancontactprob}. 
A cautionary note is in place. Unlike the contact probability exponent  calculated for each macromolecular structure, these mean properties obtained by averaging over each ensemble of proteins and RNA are meant for understanding the general difference between RNA and proteins as two distinct classes of macromolecules.

\section*{RESULTS}
\noindent{\bf Power-law exponent $\gamma$ of contact probability.}
The contact probability $P(s)$ calculated for individual biopolymers (Eq.\ref{Ps}) exhibit power-law decay over the intermediate range of $s$, $10\lesssim s\lesssim \mathcal{O}(10^2)$ (the left panel of Figs.\ref{structures}B and \ref{structures}C). 
The scaling exponent $\gamma$ from the fit using $P(s)\sim s^{-\gamma}$ was obtained for each biopolymer (see text and Figs. S1--S4 in the Supporting Material for details, where we discussed the accuracy of obtaining $\gamma$ and showed the error bar of $\gamma$ for each macromolecule), and its distributions, $p(\gamma)$s, for RNA and proteins are contrasted in Fig. \ref{structures}A. 
Proteins have $p(\gamma)$ broadly distributed from 0.5 to 2.5 centered around $\gamma \approx 1.5$, whereas $p(\gamma)$ for RNA is sharply peaked at $\gamma\approx 1.1$.  
No clear correlation is found between $\gamma$ and the chain length ($N$) in proteins; 
however, in RNA while $\gamma$ values are broadly distributed at small $N$, 
they are sharply centered around $\gamma\approx 1.1$ when $N\gtrsim 100$ (see also Fig. S6).

The distinct scaling exponents, $\gamma=1.11$ for the $P(s)$ of 23S rRNA ($P(s)$ at the left corner of Fig. \ref{structures}B) and $\gamma=1.49$ for FhuA ($P(s)$ at the left corner of Fig. \ref{structures}C), elicit special attention. 
The value of $\gamma\approx 1.0$, especially for large sized RNA arises from their characteristic chain organization: Similar to TADs in chromosomes, proximal sequences along the chain are stabilized by base pairing to form independently stable modular ``contact domains", consisting of hairpin, bulges, and loops. Further assemblies among these contact domains are achieved by a number of tertiary interactions (base triples, kissing loops, coaxial stackings through ribose zipper, A-minor motif, and metal-ion interactions) \cite{batey1999Angew,Nissen2001PNAS}. 
The abundance of distal contacts resulting from the hierarchical chain assembly likely contributes to the greater  frequency of the long-range contacts, giving rise to $\gamma\approx 1.11$ for 23S rRNA on the scale of $10\lesssim s\lesssim 300$ (see the next section).  
The distinct chain organizations of RNA and proteins become more evident when molecules are visualized  
using rainbow coloring scheme spanning the chain (Figs.\ref{structures}B and \ref{structures}C). 
The overall chain topology of 23S rRNA resembles a crumpled globule \cite{2006Luae45,Halverson14RPP} that retains clearly demarcated contact domains held by distal inter-domain contacts. 
The territorial organization of contact domains made of proximal sequences is highlighted in large sized RNA structures (see the large and small subunit of rRNA in Fig. \ref{structures}B).

In stark contrast to rRNA, typical proteins with $\gamma\approx 1.5$ (indexed with black labels from 1 to 5 in Figs.\ref{structures}A and \ref{structures}C) retain chain conformations whose subchains look topologically more intermingled with the rest of the structure, lacking visually distinct domains of a similar color. 
The intermingled chain configurations of native proteins as well as the contact probability scaling exponent $\gamma\approx 1.5$ points to a configuration of equilibrium globule, which is also supported by the same conclusion reached by investigating the loop size distribution of native protein structures \cite{2000Berezovsky283}. 
Of particular note are the proteins with $\gamma < 1.0$, which are found at the outliers of $p(\gamma)$. 
For example, $\gamma=0.73$ is for chrondroitin sulfate ABC lyase I (the protein indexed with 6) \cite{Huang2003JMB}, the chain configuration of which has clearly demarcated contact domains.

Instead of calculating the $s$-dependent contact probability for individual molecules ($P_{\alpha}(s)$, $\alpha=1,2,\ldots I_{\text{max}}$), 
one can also consider ensemble averaged characteristics of native RNA and protein organizations, $\langle\overline{P}(s)\rangle$ (Eq.\ref{meancontactprob}, Fig.\ref{structures} D).  
The mean contact probability calculated for each ensemble of RNA and proteins exhibits power-law decay $\langle \overline{P}(s)\rangle$ with  
$\gamma \approx 1.1$ for RNA and $\gamma \approx 1.6$ for proteins on the scale of $(20-30) \lesssim s \lesssim  100$, which helps us understanding the general difference of structural ensemble between RNA and proteins as two distinct classes of macromolecules.

Cautionary remarks are in place in regard to the power-law scaling of $P(s)$. 
The characteristic power-law decay behavior of 23S rRNA with $\gamma=1.1$ is only valid for intermediate range of $s$. 
For small $s$, $P(s)$ decays with a different power-law exponent (see the two panels of $P(s)$ in Figs. \ref{structures}B and \ref{structures}C).    
As reported by Lua and Grosberg \cite{2006Luae45}, on local scales both RNA and proteins have a chain organization different from the one on larger scale, which is also confirmed in our study by the distinct scaling exponent $\gamma \approx 0.4$ for RNA and $\gamma \approx 1.4$ for proteins with $s < 20$ (Fig. S7).  
Hence, in the strict sense the chain organizations of both RNA and proteins are \emph{not} scale-invariant, which is not the case for any real polymer, either.   
Depending on the length scale of interest, different picture is revealed from real polymer chains. 
Of note, the new scaling exponent $\gamma=0.75$ recently discovered for chromatin organization at resolution ($10\text{ kb}\lesssim s<1$ Mb) \cite{sanborn2015PNAS} higher than the previous study ($s\gtrsim 700$ kb) \cite{lieberman09Science} implies that the self-similarity found at the intermediate resolution ($P(s)\sim s^{-1.08}$) cannot be extended to the internal structure of contact domain with $P(s)\sim s^{-0.75}$. 
\\

\begin{figure*}[ht]
\begin{center}
\centerline{\includegraphics[width=1.7\columnwidth]{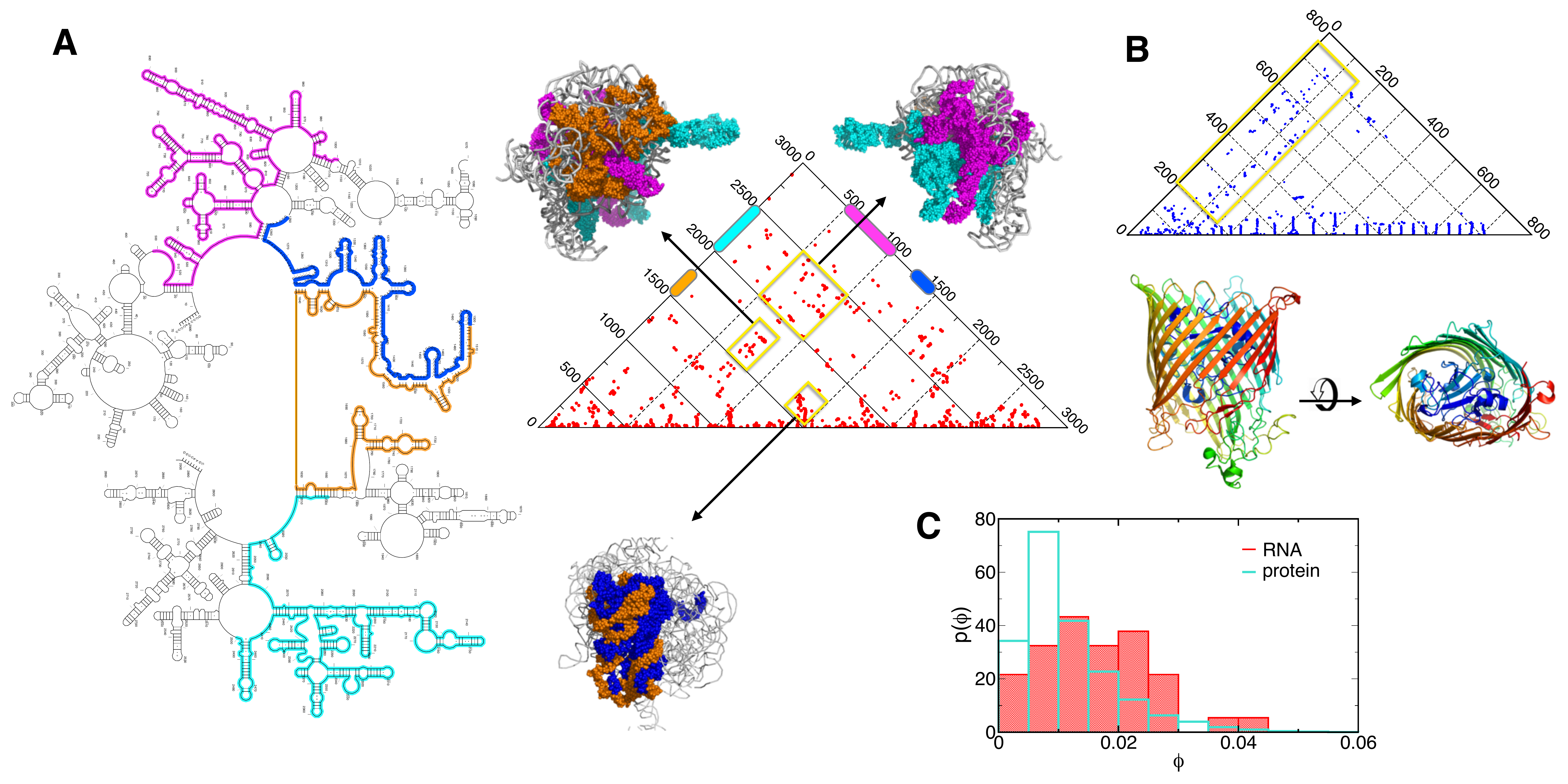}}
\caption{
Analysis of long-range contacts. 
{\bf A}, {\bf B}. Contact maps of 23S rRNA and FhuA, whose $P(s)$ are provided in Fig. \ref{structures}. 
In FhuA, the long-range contacts ($s\gtrsim 100$), enclosed by a yellow box, are formed between the structure made of N-terminal sequences ($i=1-150$) and surrounding $\beta$-strands forming the barrel. 
In 23S-rRNA, the locations of the clusters of long-range contacts formed at the interfaces between contact domains are highlighted using different colors on each range of sequences along with the three dimensional structures. 
{\bf C}. Histogram of the density of long-range contacts calculated for RNA and protein structures in the PDB.
} 
\label{contacts}
\end{center}
\end{figure*}

\noindent{\bf Long range contacts from contact map. }
Contact maps along with the three dimensional structure offer a more concrete insight into the distinct chain organization of biopolymers with different $\gamma$. 
For instance, the contact maps of 23S-rRNA ($\gamma=1.11$) (Fig. \ref{contacts}A) and FhuA ($\gamma=1.49$) (Fig. \ref{contacts}B) reveal that 23S rRNA has a greater density of long-range contacts than FhuA. 
Interestingly, in 23S rRNA the modular contact domains made of sequences, spanning $i=500-1000$ (magenta) and $1500-1750$ (orange) or between $i=500-1000$ (magenta) and $2000-2500$ (cyan), form extensive interfaces  (Fig. \ref{contacts}A). 
In comparison, FhuA has $\beta$-barrel structure with the long-range tertiary contacts formed between the subdomain (blue) made of N-terminal sequences ($i=1-150$) and $\beta$-strands ($i=200-700$) surrounding it (Fig. \ref{contacts}B).

To generalize this finding for RNA and proteins,  
for each structure we calculated the proportion of long-range contacts ($\phi$), between any sites $i$ and $j$, satisfying $j-i\ge s_{\text{min}}$, as the ratio between the observed number of long-range contacts and the maximum possible number of long-range contacts, i.e., $\phi=\mathcal{N}\sum_{j-i\ge s_{\text{min}}}^N\Theta(d_c-|\vec{r}_i-\vec{r}_j|)$, where $\Theta(\cdots)$ is the Heaviside step function and the normalization constant $\mathcal{N}=(N-s_{\text{min}}+1)(N-s_{\text{min}})/2$. 
The corresponding histograms $p(\phi)$ for RNA and proteins are shown in Fig. \ref{contacts}C with $s_{\text{min}}=30$. 
The finding that RNA has $p(\phi)$ distributed to larger $\phi$ values than proteins indicates that a significant number of tertiary contacts are used for assembling the secondary structure elements abundant in RNA. 
This result is robust to the variation of $s_{\text{min}}$ value. 
\\

\begin{figure}[h!]
\centering{\includegraphics[width=1.0\columnwidth]{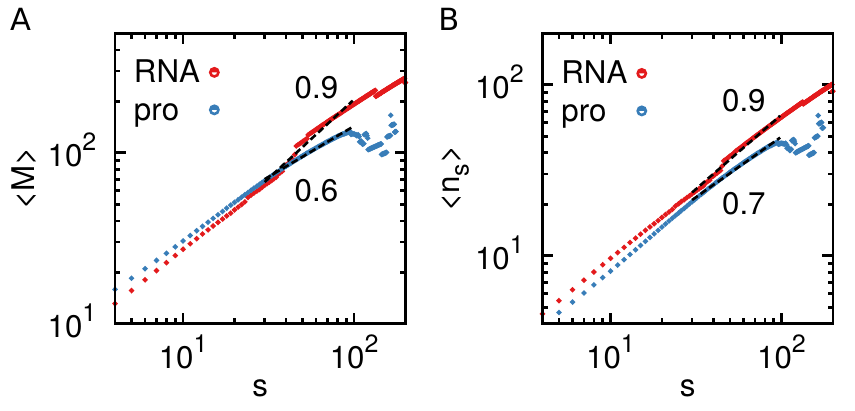}}
\caption{
Chain size frequency weighted number of inter-subchain interactions $M$ ({\bf A}) and the number of surface monomers $n_{s}$ ({\bf B}) as a function of subchain size $s$ in RNA (red) and proteins  (blue).
}
\label{subchain}
\end{figure}

\noindent{\bf Inter-subchain interactions and surface roughness. } 
In order to quantify further the distinct chain organization of RNA and proteins, we borrow analytic tools developed in the studies of chromosome organization \cite{2006Luae45,Halverson14RPP}.
The number of contacts, $M(s)$, that a subchain has with the rest of the structure (see Fig. \ref{subchain}A) \cite{Mirny11ChromoRes}, scales as  $\langle M(s)\rangle\sim s^{\beta_{1}}$ for both RNA and proteins, where $\langle\ldots\rangle$ denotes an average over the chain size frequency (see Materials and Methods). 
The exponent $\beta_1$ is different for RNA ($\beta_1^{\text{RNA}}=0.9$) and protein ($\beta_1^{\text{prot}}=0.6$), and $\langle M\rangle$ is greater for RNA when $s\gtrsim 40$, indicating that RNA has more number of inter-subchain contacts for $s\gtrsim 40$. 
The same conclusion was drawn by computing the ``roughness" of the subchain surface \cite{Halverson14RPP}, which is quantified using $n_{\text{s}}(s)$, the number of monomers in a subchain that are in contact with at least one monomer belonging to other subchains (see Fig. \ref{subchain}B). 
$\langle n_{s}(s)\rangle_L \sim s^{\beta_{2}}$ with $\beta_2^{\text{RNA}}=0.9>\beta_2^{\text{prot}}=0.7$, suggesting that RNA have rougher subchain surfaces.
The scaling relationships of the inter-subchain interactions ($M\sim s^{0.9}$) and the surface monomers ($n_s\sim s^{0.9}$) for RNA compare well with those of crumpled globules ($M$, $n_s\sim s^1$) \cite{Mirny11ChromoRes,Halverson14RPP}.  

$\langle M(s)\rangle$ and $\langle n_s(s)\rangle$ are related to each other with $\langle M(s)\rangle\approx Q\langle n_s(s)\rangle$ where $Q\sim s^{\nu d}/s$ is the proportionality constant, the total number of possible monomers ($\sim s^{\nu d}$) that can fill the volume defined by a blob consisting of $s$ monomers, thus giving a scaling relation 
$\beta_1=\nu d-1+\beta_2$ \cite{Halverson14RPP}. 
From this relation and $\beta_{1,2}$, we obtain the Flory exponent $\nu=1/3$ for native RNA and 
$\nu=0.3$ for proteins, which is in perfect agreement with the values of $\nu$ obtained from an independent analysis of macromolecular structures in the PDB, $\nu\approx 0.33$ for RNA and $\nu=0.31$ for proteins in $R_G\sim N^{\nu}$ \cite{Hyeon06JCP_2}. 
\\

\begin{figure}[h!]
\centering{\includegraphics[width=1.0\columnwidth]{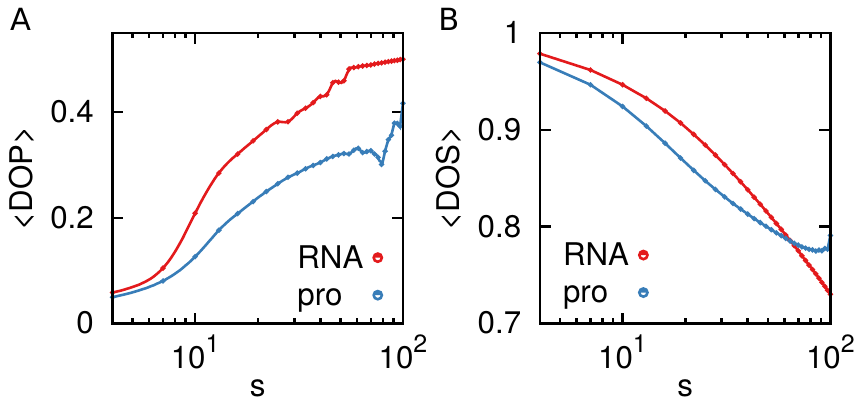}}\vspace*{-5.5pt}
\caption{
The mean degree of interpenetration (DOP) and segregation (DOS) as a function of subchain size $s$ in RNA (red) and in proteins (blue).
}
\label{DOSDOP}
\end{figure}  

\noindent{\bf Degree of interpenetration and segregation. } 
Next, we calculate the fraction of residues from other subchains found in the ellipsoidal volume enclosing a subchain averaged over all subchains of length $s$, which corresponds to the degree of interpenetration (DOP) \cite{2006Luae45}. 
The degree of segregation (DOS), $\text{DOS}=\langle d_{A,B}/(2 R_G^{A\cup B})\rangle$ is defined by the ratio between  $d_{A,B}$ and $(2 R_G^{A\cup B})$, 
where $d_{A,B}$ is the distance between the center positions of two non-overlapping subchains $A$ and $B$, 
and $R_G^{A\cup B}$ is the gyration radius of the union of these two subchains. 
DOS is defined by the ratio of these two values ($d_{A,B}$ and $R_G^{A\cup B}$) averaged over all the pairs of subchains $A$ and $B$ with the same length $s$. 
DOP and DOS as a function of $s$ for both RNA and proteins (Fig. \ref{DOSDOP}) indicate that while subchains separated by a large arc length $s$ are well separated from each other in RNA 
the subchains in RNA penetrate into other subchain's volume deeper than in proteins. 
This explains why the decline of $P(s)$ for RNA is slower than for proteins (Fig.  2), which leads to a smaller exponent, $\gamma$.  
\\

\begin{figure}[h!]
\centering{\includegraphics[width=1.0\columnwidth]{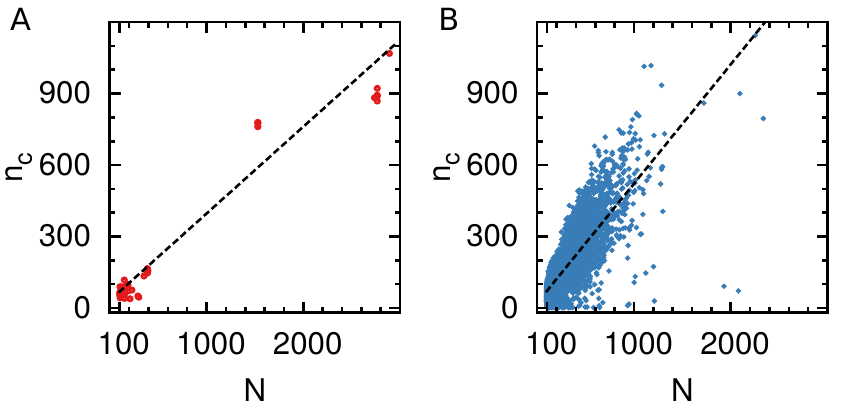}}
\caption{
Scatter plots of the number of contacts  for RNA (left) and proteins (right) over the intermediate range with $s_{\text{min}}=30$ and $s_{\text{max}}=100$.
}
\label{NcL}
\end{figure}

\noindent{\bf The number of long-range contacts. }
The total number of contacts over a given range of $s$, $s_{\text{min}}<s<s_{\text{max}}$ is considered with $P(s)\approx q s^{-\gamma}$:   
$n_c(N)= \int_{s_{\text{min}}}^{s_{\text{max}}} (N-s) P(s) ds \approx q\int_{s_{\text{min}}}^{s_{\text{max}}} (N-s) s^{-\gamma} ds$, and hence 
\begin{align}
n_c(N)/q\approx N\left(\frac{s_{\text{max}}^{1-\gamma}-s_{\text{min}}^{1-\gamma}}{1-\gamma}\right)+\left(\frac{s_{\text{max}}^{2-\gamma}-s_{\text{min}}^{2-\gamma}}{2-\gamma}\right).
\label{nc}
\end{align}
Notably,  
(i) $n_c/q$ scales linearly with $N$ for both RNA and proteins, regardless of $\gamma$ value.  
(ii) The prefactor of $n_c/q$ depends only on $\gamma$.
For $s_{\text{min}}=30$ and $s_{\text{max}}=100$, Eq.\ref{nc} leads to  
$n_c^{\gamma=1.1}(N)/q\approx 0.81 N+46$, and 
$n_c^{\gamma=1.6}(N)/q\approx 0.11 N+6.0$. 

Meanwhile, from the plots of $n_c(N)$ using structures in PDB (see Fig. \ref{NcL}), we obtain  
\begin{align}
n_c^{\text{RNA}}(N)/q^{\text{RNA}}&\approx  0.77 N + 65\nonumber\\
n_c^{\text{pro}}(N)/q^{\text{pro}} 
&\approx 0.11 N + 4.7, 
\end{align}
where the prefactors $q^{\text{RNA}}\approx 0.48$ and $q^{\text{pro}}\approx 4.71$ from the fits to $\langle\overline{P}(s)\rangle$ in Fig. \ref{structures} are used. 
Note that for a given $N$, $n_c^{\gamma=1.1}(N)>n_c^{\gamma=1.6}(N)$ and $n_c^{\text{RNA}}(N)>n_c^{\text{pro}}(N)$.  
Together with other quantities, $\langle\text{DOP}\rangle$, $\langle\text{DOS}\rangle$, $\langle M(s)\rangle$ and $\langle n_s(s)\rangle$, 
the number of contacts, $n_c(N)$, calculated here persistently assert that RNA have greater number of long-range contacts than proteins of the same size.

It is of note that the analyses presented in Figs. 3, 4, 5 are different from investigating each macromolecule one by one (Fig. 1) and finding the structure-function relationship. Given that the ensemble in question is the product of evolution, clarifying the difference between two classes of macromolecules (RNA and proteins) is promising as soon as the evolutionary questions are concerned.

\section*{DISCUSSION}
Due to intramolecular forces stabilizing chain molecule, both native RNA and protein molecules retain compact and space-filling structures, satisfying $R_{G}\sim N^{1/3}$ \cite{Florybook,Hyeon06JCP_2}, which from the polymer physics perspective is regarded as the property of polymers in poor solvent conditions.     
It is, however, critical to note that the size of a \emph{subchain} surrounded by other subchains should scale as $R(s)\sim s^{1/2}$, which is indeed confirmed for the proteins with $\gamma=1.5$ (Fig.S5). 
According to the ``Flory theorem" \cite{Flory49JCP,GrosbergBook}, a test chain in a \emph{fully} equilibrated homogeneous semi-dilute  or concentrated polymer melt \cite{deGennesbook},  in spherical confinement \cite{Cacciuto2006NanoLett}, or even in globule is expected to obey the Gaussian statistics because of the screening of excluded volume interaction or counterbalance between attraction and repulsion \cite{GrosbergBook}, thus satisfying $R(s)\sim s^{1/2}$ or $P(s)\sim s^{-3/2}$ (see Supporting Material).
The distinct contact probability exponent is highlighted by our analysis that $\gamma \sim 1.0$ for large RNA and $\gamma \sim 1.5$ for small RNA or globular proteins over the intermediate range of $20\lesssim s\lesssim 100$. 
Evident from rRNA structure (Fig. \ref{structures}), 
subchains of RNA at scales $s>20$ are assembled into modular contact domains, which are better demarcated in the form of stem-loop helices than proteins, and stitched together through long-range tertiary contacts (Fig. \ref{contacts}A). 
The evidence of this characteristic architecture of RNA with multi-modular domains is visualized vividly in the form of multiple rupture events in single molecule pulling experiments of \emph{T}. ribozymes \cite{Onoa2003Science}, while many proteins display a cooperative and effectively all-or-none unfolding under force \cite{shank2010Nature,Mickler07PNAS}.

What causes the crumpled structures of large RNA at the scale of  $20\lesssim s\lesssim 100$?  
Here, 
the statistical rarity of knots in native RNA \cite{2015Micheletti2052,burton2015RNAbiol}, which is unparalleled by proteins or DNA  \cite{2006Luae45,noel2010PNAS}, is worth noting. 
In general, knots are unavoidable when a long polymer chain ($N\gg N_e\sim 200$ where $N_e$ is the entanglement length \cite{deGennesbook}) 
is folded to an equilibrium globule \cite{grosberg2000PRL,Mirny11ChromoRes}. 
Topological knot-free constraints inherent to the ring polymers, however, have been shown to organize melts of unconcatenated polymer rings or a single long polymer ring into crumpled globules, preventing entanglements \cite{imakaev2015SoftMatter,Halverson14RPP}. 
Since large RNA molecules, assembled by a number of secondary structural elements (hairpin loops, stems), resemble a collection of small and large rings, it can be surmised during the folding process, the knot-free constraints are effectively imposed. 
The knot-free constraints are more likely applied for RNA because the energy scale associated with secondary structure elements ($\varepsilon^{\text{sec}}$), is in general well separated from that of tertiary interactions ($\varepsilon^{\text{ter}}$), such that $\sum_{i}\varepsilon^{\text{sec}}_i\gg \sum_k\varepsilon^{\text{ter}}_k\gg k_BT$ \cite{Thirumalai08Noncoding}, which makes secondary structure elements \emph{independently stable}.   
By contrast, in order to fold, proteins undergo reptation-like process, after the initial collapse \cite{Thirum95JPI}, which may take place with ease because secondary structure elements of proteins ($\alpha$-helix, $\beta$-sheet) are only marginally stable relative to the thermal energy. If necessary, these motifs can be reassembled into thermodynamically more stable structures. 

While local and remote contacts are mixed in the folding nuclei of proteins, 
the formation of secondary structures in RNA folding usually precedes the formation of tertiary contacts, so that 
the folding of RNA is hierarchical \cite{Tinoco99JMB,Greenleaf08Science}. 
Folding under kinetic control produces thermodynamically metastable and kinetically trapped intermediates, which occurs ubiquitously in RNA folding \cite{Treiber01COSB}, especially in cotranscriptional folding of RNA \cite{Repsilber1999RNA,Lutz2013NAR}.  
A decision, made at an early stage of folding, involved with the formation of independently stable secondary structure elements is difficult to revert although in a worst case scenario,  cofactors such as metal-ions \cite{Wu98PNAS,Koculi2012NAR}, metabolites \cite{Montange08ARB}, and RNA chaperones \cite{Russell2013RNAbiology} still can induce a secondary structure rearrangement. 
Hence, a more proper way to understand conformational dynamics of a large RNA molecule with $N\gtrsim 100$ is to consider an ensemble of multiple functional states \cite{AlHashimi08COSB,Solomatin10Nature,Hyeon2012NatureChem,Hyeon14PRL} instead of a thermodynamically driven, unique native state.
It is noteworthy that RNA secondary structure prediction algorithms which use the strategy of searching the minimum free energy structure \cite{Rivas99JMB,hofacker03NAR,Zuker03NAR} fail to predict the correct secondary structure when $N\gtrsim 100$, and require the comparative sequence analysis or experimental constraints \cite{GutellCOSB02,MathewsJMB99}.   
This could be ascribed to the consequence of error accumulated in predicting RNA structures with large $N$, but it is also suspected that the (free) energy minimization principle cannot be extended to account for the folding process of large RNA. 
The contact statistics of large RNA, $P(s) \sim s^{-1}$, 
can be used as an additional constraint or guideline for structure prediction.

A situation analogous to the hierarchical folding of large RNA is prevalent in the two-stage membrane protein folding where the insertion of transmembrane (TM) $\alpha$-helices, guided by translocons, is followed by the post-insertion folding  \cite{Popot1987JMB,Bowie05Nature}. 
We indeed find that the contact probabilities of class A G-protein coupled receptors (GPCRs) give $\gamma\approx 1$ (blue circles in the middle panel of Fig. \ref{structures}). 
Since $\gamma\approx 1$ means the chain organization of native GPCRs is not in entropy-maximum state, a thermodynamically guided, spontaneous \emph{in vitro} refolding of GPCRs into the native form is expected to be non-permissible.   
An AFM experiment on an $\alpha$-helical membrane protein, antiporter ($N\approx 380$), whose $\gamma$ value we find is $\approx 1.1$, could not be refolded to the original form after mechanically unfolded \cite{kedrov2004JMB}.  
However, a recent remarkable single molecule force experiment \cite{Min15NCB} has shown that GlpG, an $\alpha$-membrane proteins with $N\approx 270$, can reversibly fold in bicelles even after the entire structure including TM helices is disrupted by mechanical forces. 
Remarkably, we find $\gamma\approx 1.5$ for GlpG. 
For membrane proteins of known native structures, 
their $\gamma$ values can be used to judge whether or not spontaneous \emph{in vitro} refolding is possible.


Since the time required for equilibrium sampling of conformations ($\tau_{eq}$) increases exponentially with the system size ($N$) as $\tau_{eq}\sim e^N$ \cite{Palmer82AP}, signatures of metastability or non-equilibration in chain conformation could be ubiquitous in a macromolecular structure with large $N$. 
Through the statistical analysis of structures in PDB, our study puts forward that 
the present forms of crumpled chain organization with $\gamma\approx 1$ of large native RNA and some classes of proteins are an ineluctable outcome of the folding mechanism under kinetic control. 

Our results, based on the structures available in PDB, might be fraught with a possible sample bias since the current structural information available in PDB is limited underrepresenting intrinsically disordered proteins or membrane proteins for proteins, and long non-coding intron RNA (lncRNA) abundant in the cell for RNA \cite{rinn2012ARBiochem,carninci2005Science}. Nevertheless, our general conclusions on the difference in the organization principle between proteins and RNA will still hold even when the database of PDB is further expanded. 
Especially, we expect that an inclusion of long non-coding RNA structures ($N > 200$), which should be possible in the near future, will make our conclusions more robust since the hierarchical nature of RNA folding process would become more evident for RNA with larger $N$ and reinforce the territorial (crumpled-like) organization in RNA.
\\

\section*{ACKNOWLEDGMENTS}
We thank the Korea Institute for Advanced Study for providing computing resources (KIAS Center for Advanced Computation, Linux Cluster System) for this work.


\clearpage 

\setcounter{figure}{0}  
\setcounter{equation}{0}
\makeatletter 
\renewcommand{\thefigure}{S\@arabic\c@figure}
\renewcommand{\theequation}{S\@arabic\c@equation}
\makeatother 

\section{Supporting Material}
{\bf Extraction of Scaling Exponent $\gamma$} 
We obtained the contact probability exponent $\gamma$ by conducting linear regression on a part of $P(s)$ data that behave as $\sim s^{-\gamma}$ in log-log scale. 
There are two factors that may affect in determination of $\gamma$: 
(i) $d_{c}$, the cut-off distance to define a contact between two residues, affects the overall shape of $P(s)$; 
(ii) The range of $s$, $s_{\text{min}}<s<s_{\text{max}}$, to be fitted.  
Instead of manually tuning the fitting range ($s_{\text{min}}<s<s_{\text{max}}$), 
we defined a parameter $\varphi$ ($0<\varphi<1$), such that the proportion of fitting range, $(s_{\text{max}}-s_{\text{min}})/N$ where $N$ is the chain length, is at least greater than an allocated threshold value, $\varphi$. 
For instance, if $\varphi$ is set to 0.3 then the fit is made on more than 30 \% of the entire data points.   
Thus, by fitting $P(s)$ data over all possible pairs of $s_{\text{min}}$ and $s_{\text{max}}$ values which define the range of $(s_{\text{min}}, s_{\text{max}})$ satisfying  
$(s_{\text{max}}-s_{\text{min}})/N \geq \varphi$, we determine the value of $\gamma$ from the best fit which gives the smallest standard error relative to the data points.

\begin{figure}[h!]
\begin{center}
\includegraphics[width=0.8\columnwidth]{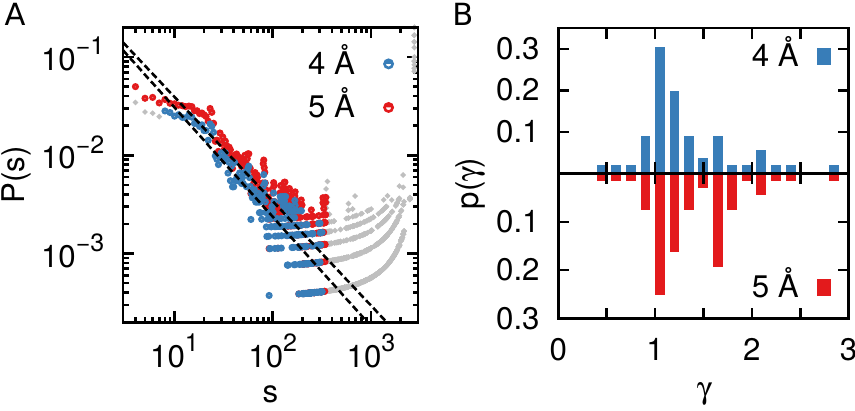}
\end{center}
\caption{
(A) The contact probability versus sequence distance of 23S rRNA (PDB entry 2O45) 
with a cut-off distance of contacting $d_{c}$ of value 4 \AA\ (blue) and 5 \AA\ (red). 
(B) Distributions of $\gamma$ in RNA monomers with $d_{c}$ of 4 \AA\ and 5 \AA.}
\label{g2dc}
\end{figure}

Fig. \ref{g2dc}A shows that the shape of $P(s)$ for 23S-rRNA calculated with different $d_{c}$ remains effectively identical, giving rise to a similar value of $\gamma$: $\gamma=1.11$ ($d_c=4$ \AA), 1.06 ($d_{c}=5$ \AA). 
$p(\gamma)$s for RNA molecules obtained from different $d_{c}$ are also similar as shown in Fig. \ref{g2dc}B.

\begin{figure}[ht]
\begin{center}
\includegraphics[width=0.8\columnwidth]{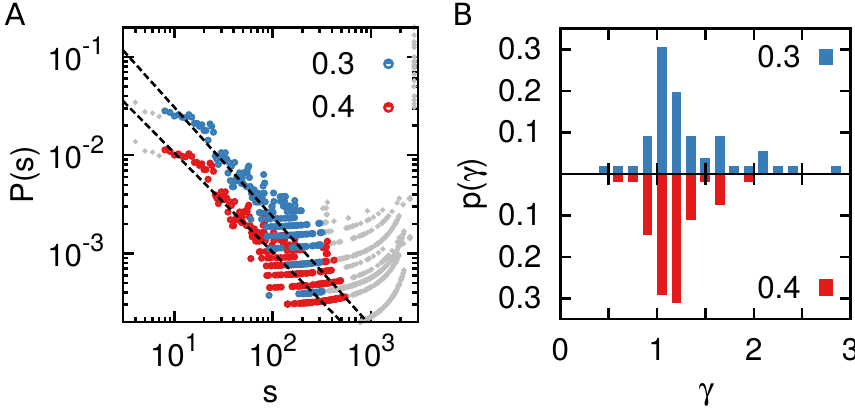}
\end{center}
\caption{
(A) The contact probability versus sequence distance of 23S rRNA (PDB entry 2O45) 
with a minimum fraction of all data points used for fitting $\varphi$ of 0.3 (blue) and 0.4 (red). 
The data points for $\varphi=0.4$, as well as the fitted dashed line, are shifted downwards for visual comparison. 
(B) Distributions of $\gamma$ in RNA monomers of $\varphi$ 0.3 and 0.4. }
\label{g2sf}
\end{figure}

Next, to study the effect of $\varphi$ on $\gamma$, we set $d_{c}=4$ \AA\ and change the value of $\varphi$ in the fit. 
We obtain  
$\gamma=1.11$ for $\varphi=0.3$, and $\gamma=1.01$ with $\varphi=0.4$ (see Fig. \ref{g2sf}A). 
Fig. \ref{g2sf}B also shows that $p(\gamma)$ with different $\varphi$ are comparable. 
Analysis applied to protein shows similar results. 
A series of comparisons in Figs.\ref{g2dc} and \ref{g2sf} indicate that the average value of $\gamma$ is insensitive to the parameters around the value we have chosen. 

\begin{figure}[ht]
\begin{center}
\includegraphics[width=0.8\columnwidth]{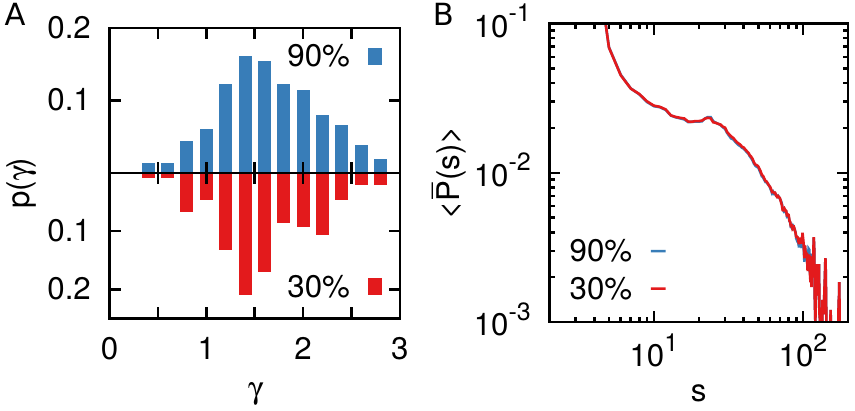}
\end{center}
\caption{Effects of imposing different threshold value for the sequence similarity of 90 \% and 30 \% to the protein structure database to compute $p(\gamma)$ and $\langle\overline{P}(s)\rangle$.  
No qualitative difference is found in the results.   
}
\label{red30}
\end{figure}

In addition, the overall shapes of $p(\gamma)$ and $\langle\overline{P}(s)\rangle$ are insensitive to the two threshold values of sequence similarity (90 and 30 \%), which we imposed to select a set of non-homologous proteins (Fig. \ref{red30}).

\begin{figure}[ht]
\begin{center}
\includegraphics[width=1\columnwidth]{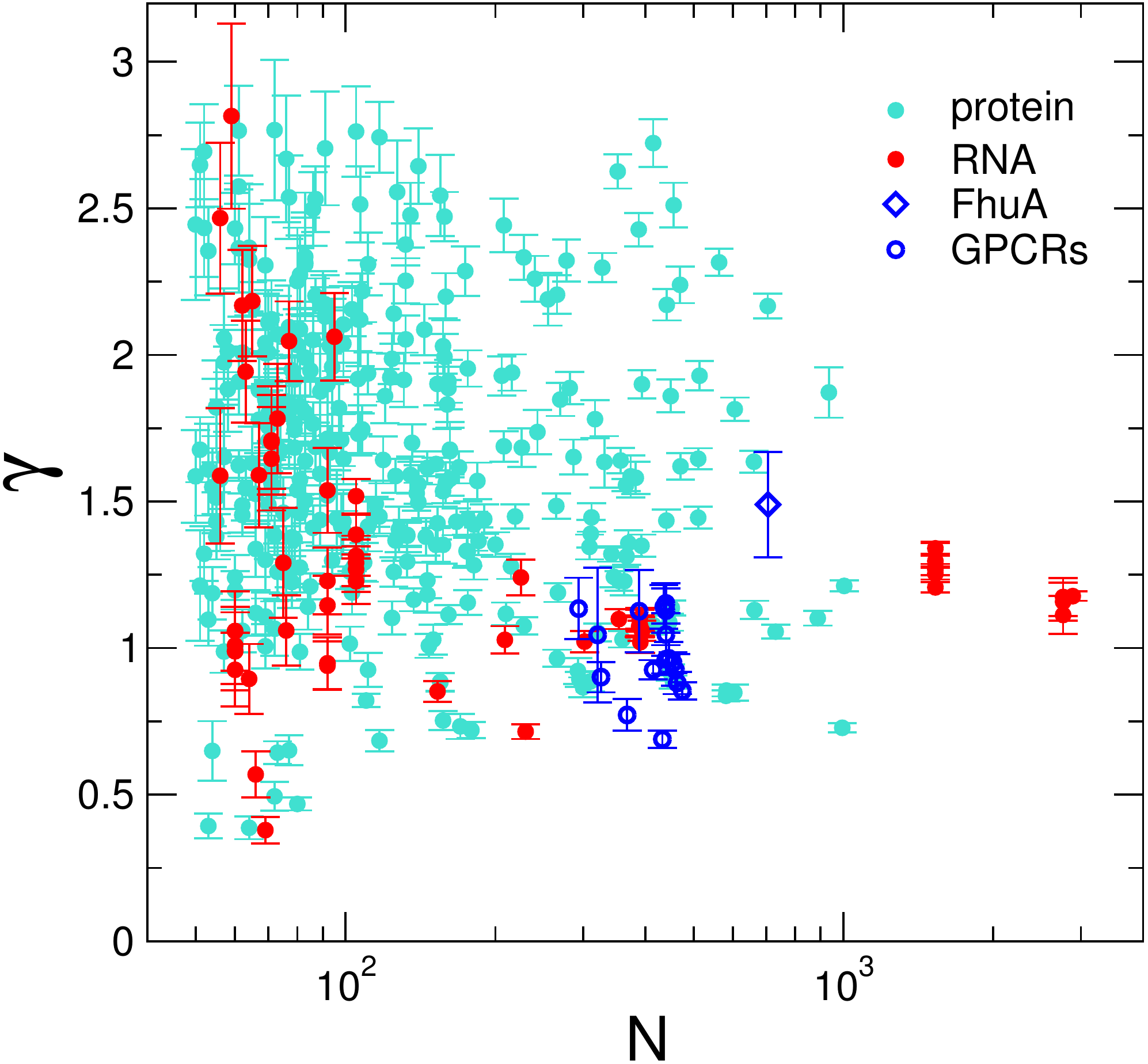}
\end{center}
\caption{
Scatter plot of $\gamma$ versus $N$ for RNA (red) and proteins (cyan) with error bars (95 \% confidence interval) for $\gamma$ values. 
}
\label{gError}
\end{figure}

We analyzed 186 RNA and 16633 individual proteins whose size satisfies $N \geq 50$, 
available in PDB as of September 2015. 
Distributions of $\gamma$ obtained from the optimal linear fittings on $\log_{10}P(s)$ versus $\log_{10}s$ with a correlation coefficient greater than 0.9 are presented in Fig.1A with $\varphi=0.3$, $d_c=4$ \AA\ for both RNAs and proteins. 
To highlight the robustness of our result presented in Fig.1A ($\gamma$ vs. $N$ plot), we specified the 95 \% confidence interval of $\gamma$ values using error-bar to each data point in Fig. \ref{gError}. 
\\

\begin{figure}[ht]
\begin{center}
\includegraphics[width=0.8\columnwidth]{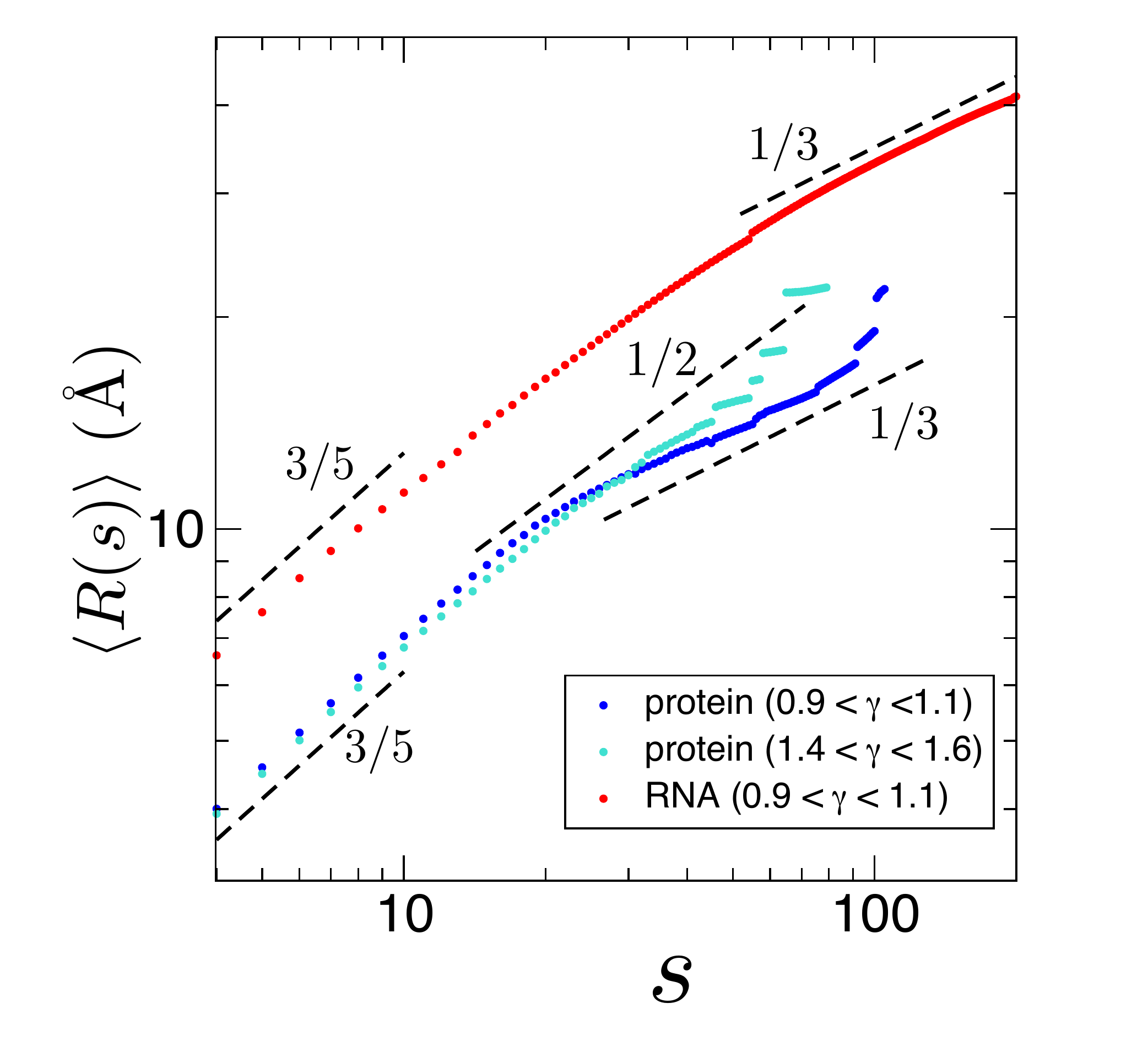}
\end{center}
\caption{Mean radius of gyration of subchain as a function of subchain length $s$ for proteins and RNA that display contact probability exponent in a specified range of $\gamma$. The structures in the specified range of $\gamma$ were collected from Fig. 1 and their $R(s)$s were calculated.}
\label{rs}
\end{figure}

{\bf Contact probability between two sites of a polymer}
In general, the contact probability of two sites in polymer chain is determined by the volume available for the subchain  ending with the two sites, $[R(s)]^d$, with normalization condition $\int P_s(r)d^dr=1$ \cite{GrosbergBook,deGennesbook}:
\begin{align}
&P_s(r)=\frac{1}{R(s)^d}\varphi\left(\frac{r}{R(s)}\right)\nonumber\\
&\xrightarrow{r\ll R(s)} \nonumber\\
&P(s)=\frac{1}{R(s)^d}\left(\frac{r}{R(s)}\right)^g, 
\end{align} 
where $r$ is the contact distance, $R(s)$ is the size of polymer made of $s$ monomers, $d$ is the dimensionality,  and $g$ is the correlation hole exponent. 
With $R(s)\sim s^{\nu}$ (see Fig. \ref{rs}), we obtain the scaling relationship of contact probability, $P(s)\sim s^{-\nu(d+g)}$. 

(i) When the excluded volume interaction is fully screened, a test chain (or subchain over a certain length) is ideal. 
In this case, $\varphi(x)\sim e^{-3x^2/2}$. 
Thus, the correlation hole exponent $g=0$ \cite{FriedmanJPII91} and $R\sim s^{\nu}$ with $\nu=1/2$, which leads to $P(s)\sim s^{-\nu d}\sim s^{-3/2}$. 

(ii) If the chain adopts an \emph{effectively homogeneous} space-filling configuration, but the interaction between monomers is weak and the excluded volume interaction is still fully screened as in a concentrated melt, then $g=0$, $d=3$, and $\nu=1/3$, which leads to $P(s)\sim s^{-1}$. 

(iii) If the chain organization is \emph{inhomogeneous} leading to an anisotropic arrangement because of strong monomer-monomer interactions \cite{sanborn2015PNAS}, which for the case of RNA leads to formation of independently stable helices, then $R(s)$ still satisfies $R(s)\sim s^{1/3}$ but the effective dimensionality of the sampling space ($d_{\text{eff}}$) would be less than $3$. 
Thus, $P(s)\sim s^{-d_{\text{eff}}/3}$, and $\gamma=d_{\text{eff}}/3<1$, which accounts for the contact probability exponent smaller than 1.   

(iv) Note that when the subchain interactions (repulsion and attraction) are screened ($g=0$), 
$P(s)$ and $R(s)$ are related as $P(s)\sim R(s)^{-d}$. 
This relationship particularly holds good for intermediate range of $s$: $P(s)\sim s^{-3/2}\leftrightarrow R(s)\sim s^{1/2}$ (ideal chain) and $P(s)\sim s^{-1}\leftrightarrow R(s)\sim s^{1/3}$ (crumpled chain) (see Fig. \ref{rs}). 
The scaling exponent of $3/5$ at $s<10$ in Fig. \ref{rs} is due to the volume exclusion interaction at short range $s$.

\begin{figure}[ht]
\centerline{\includegraphics[width=0.8\columnwidth]{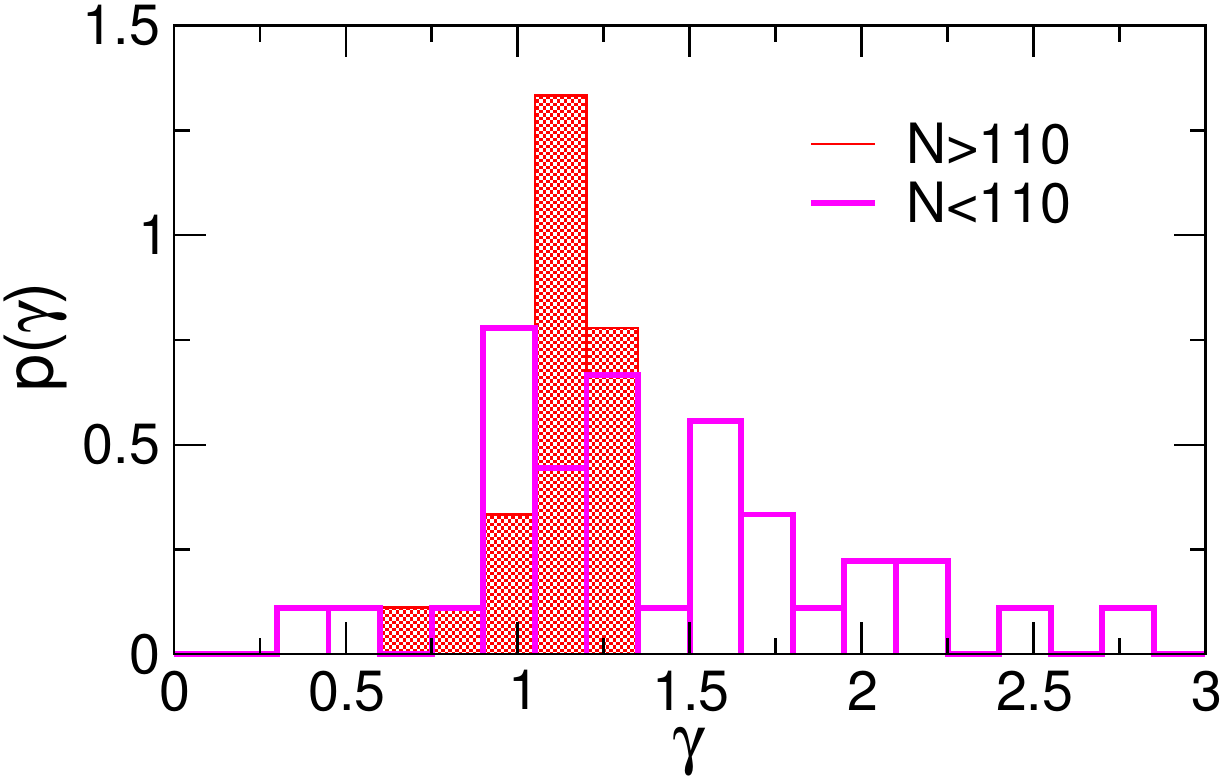}}
\caption{Distribution of $\gamma$ value for RNA with $N>110$ and $N<110$.
$\gamma^{\text{RNA}}_{N>110}=1.12\pm 0.14$ and $\gamma^{\text{RNA}}_{N<110}=1.41\pm 0.53$.
}
\label{gamma_distr_110}
\end{figure}

\begin{figure}[ht]
\centerline{\includegraphics[width=0.8\columnwidth]{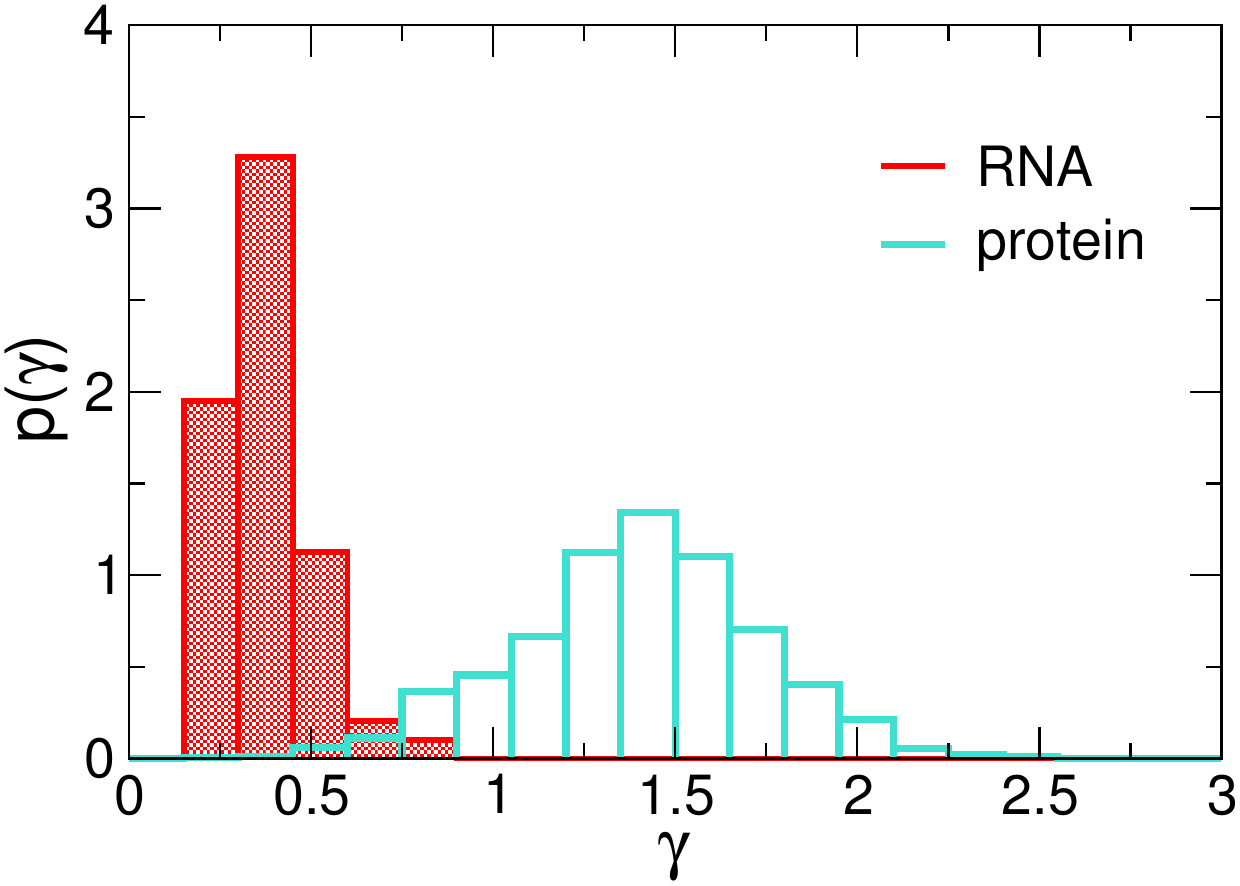}}
\caption{Distribution of contact probability exponent calculated for the short range of $s$, $s<20$.
$\gamma^{\text{RNA}}=0.38\pm 0.13$ and $\gamma^{\text{pro}}=1.40\pm 0.33$.
}
\label{gamma_distri_smin}
\end{figure}

\begingroup
\squeezetable
\begin{table}[!ht]
\centering
\tabcolsep=3pt
\begin{tabular}{@{}llllllll@{}} \hline
2M58 & 2MIY & 1FIR & 6TNA & 1EHZ & 1TRA & 4TRA & 1TN1  \\
1TN2 & 3TRA & 2TRA & 3BBV & 1VTQ & 4PQV & 3A3A & 3CW6  \\
2HOP & 1I9V & 3L0U & 2K4C & 3D2G & 4NYD & 2HOM & 3GX6  \\
2GIS & 3GX2 & 3IQN & 4B5R & 2YDH & 4RZD & 3F2Q & 3F2W  \\
3F30 & 3F2X & 3F2T & 3F2Y & 1U9S & 3DHS & 1Y0Q & 4C4Q  \\
2A2E & 3BWP & 4FAX & 4E8P & 4E8R & 4E8Q & 4E8N & 4DS6  \\
4E8M & 4FAQ & 3J2B & 3J2H & 3J2D & 2YKR & 3J28 & 3J2A  \\
2O45 & 2O43 & 2O44 & 1C2W &      &      &      &       \\ \hline
\end{tabular}
\caption{PDB entries of RNA analyzed in Fig. 1.}
\end{table}
\endgroup

\begingroup
\squeezetable
\begin{table}[!hb]
\centering
\tabcolsep=3pt
\begin{tabular}{@{}llllllll@{}} \hline
2MGW & 2JY5 & 2CR8 & 2RRU & 2KAK & 2DAH & 2EPS & 1JJR  \\
1KMX & 2KQB & 1YSM & 2ECM & 2KMU & 1KFT & 2KPI & 2M8E  \\
2K2T & 2REL & 2YSD & 2L4E & 2MWR & 2YRG & 3GOH & 1Z60  \\
2KKJ & 1A7I & 1VYX & 2M2F & 2JXD & 2DAL & 3WIT & 2M9W  \\
2YSJ & 1UEO & 1AA3 & 4A3N & 1WG2 & 2D8U & 1WFH & 1HYI  \\
1BW5 & 2DZL & 1X4P & 1VFY & 1X4W & 1HTA & 1SF0 & 1H0Z  \\
2EA6 & 2MFK & 2DI0 & 2EWT & 2RMR & 3H33 & 1RIY & 4TXA  \\
2DA7 & 2LGW & 2JVG & 1X61 & 1WEE & 1X4K & 2DJB & 4P3V  \\
2CT5 & 2LEK & 2HI3 & 1G33 & 2EP4 & 1NEQ & 1APJ & 1WFP  \\
2JXW & 2KW9 & 1SIG & 2M4G & 2LT1 & 1WYS & 1X68 & 2ENN  \\
2E6S & 2D9H & 2ECT & 1E4U & 1JQ0 & 1J3C & 1MJ4 & 4U12  \\
2MLB & 1UHC & 2CR7 & 1KDU & 1QRY & 1X3H & 2CSY & 2ECL  \\
1RWJ & 2LDR & 4CIK & 3J0R & 1UHA & 4EIF & 1X63 & 2DOE  \\
2LQL & 1CC5 & 1XFE & 2L0S & 3CP1 & 3ZJ1 & 3BT4 & 2LRQ  \\
1IPG & 2Q18 & 4IYL & 2ECW & 2LV2 & 1LMJ & 1ABA & 1C9F  \\
1F1F & 2CT2 & 1C6R & 1FP0 & 2KW1 & 4GPS & 1CTJ & 2M5W  \\
1Y02 & 2D8Y & 2E6R & 1WEO & 2CS3 & 1FBR & 2LGX & 2LGP  \\
2MIQ & 1SJ6 & 1WIA & 2JSN & 2DMD & 2VTK & 3PO8 & 1OPC  \\
2YRE & 2LGV & 1T1D & 3H6N & 1JHG & 4BGC & 2OA4 & 2CQK  \\
2CTK & 3GCE & 2K4J & 3DQY & 1X0T & 2JVL & 1HKF & 2CS8  \\
3O8V & 3DVI & 2CTW & 4EEU & 2MLK & 1ZOX & 2XXC & 2EO3  \\
4TVM & 2IVW & 2LW4 & 4HWM & 2KQR & 2JXN & 2HC5 & 1T6A  \\
4ZBH & 1UJX & 2MMZ & 2LHT & 1JUG & 2RA9 & 2XWS & 1G3P  \\
2QYZ & 2FYG & 3O5E & 2ES0 & 4NAZ & 3E2I & 1DQG & 1VSR  \\
1KQW & 1E29 & 2FVV & 3W9K & 1NL1 & 1WK0 & 1XN5 & 2IN0  \\
2NWF & 2L5Q & 2P0B & 2MO5 & 3ZUI & 2HNA & 2JY9 & 4MYM  \\
3N9D & 2N48 & 4M4Z & 3FME & 1ENV & 2D37 & 2XB3 & 1ZND  \\
4GNY & 4LD1 & 3UF4 & 1D7P & 1EW3 & 3OUQ & 1E88 & 2LFU  \\
2KIG & 2KFU & 1KLO & 3NZM & 2M47 & 4JHG & 1RL6 & 3TXO  \\
2LZM & 2NN5 & 3W9R & 2CP6 & 4F47 & 1EH6 & 1CDY & 2R6V  \\
3K21 & 3WJT & 1WV3 & 4M6T & 2D5M & 3KBG & 1J3G & 1EJE  \\
1JM1 & 3TFM & 4QA8 & 1HXN & 4E1B & 4IT3 & 4JZC & 1EMA  \\
2K18 & 3HBK & 3NO3 & 4PQ0 & 2PNN & 1LVA & 3LTI & 4JS8  \\
4DWO & 2A1L & 4NW4 & 3V75 & 5BN7 & 1DUW & 3JRP & 2QLU  \\
2LQW & 4X36 & 2HES & 4GGC & 4GGA & 4V16 & 4AA8 & 2FGQ  \\
4AF8 & 1VPR & 2PMN & 2XE1 & 2ASI & 2ZYL & 1T6E & 3BA0  \\
1J6Z & 4QDC & 4GQ1 & 1FEP & 3GRE & 4UQE & 4MSX & 3R1K  \\
3ACP & 2DH2 & 4COT & 3DWO & 1QCF & 1FMK & 1W52 & 1DQ3  \\
1G0D & 3K5W & 2OBD & 4NOX & 4FWW & 2E84 & 1Z1N & 4AW7  \\
1XEZ & 4TLW & 1PI6 & 4UMW & 4BBJ & 3OKT & 1QFG & 4MHC  \\
2OAJ & 4UP5 & 1HN0 & 3KLK &      &      &      &       \\ \hline
\end{tabular}
\caption{PDB entries of proteins analyzed in Fig. 1.}
\end{table}
\endgroup

\begingroup
\squeezetable
\begin{table}[!h]
\centering
\tabcolsep=3pt
\begin{tabular}{@{}llllllll@{}} \hline
2JX9 & 1ISR & 2LNL & 2RH1 & 2YDV & 2ZIY & 3C9L & 3EHS  \\
3EML & 3N94 & 3OE6 & 3RZE & 3UON & 3V2Y & 3VW7 & 4DKL  \\
4EJ4 & 4F11 & 4IB4 & 1QJQ &      &      &      &       \\ \hline
\end{tabular}
\caption{PDB entries of GPCRs analyzed in Fig. 1.}
\end{table}
\endgroup


\begin{thebibliography}{75}
\expandafter\ifx\csname natexlab\endcsname\relax\def\natexlab#1{#1}\fi
\expandafter\ifx\csname bibnamefont\endcsname\relax
  \def\bibnamefont#1{#1}\fi
\expandafter\ifx\csname bibfnamefont\endcsname\relax
  \def\bibfnamefont#1{#1}\fi
\expandafter\ifx\csname citenamefont\endcsname\relax
  \def\citenamefont#1{#1}\fi
\expandafter\ifx\csname url\endcsname\relax
  \def\url#1{\texttt{#1}}\fi
\expandafter\ifx\csname urlprefix\endcsname\relax\def\urlprefix{URL }\fi
\providecommand{\bibinfo}[2]{#2}
\providecommand{\eprint}[2][]{\url{#2}}

\bibitem[{\citenamefont{Schuster et~al.}(1994)\citenamefont{Schuster, Fontana,
  Stadler, and Hofacker}}]{schuster1994PRSL}
\bibinfo{author}{\bibfnamefont{P.}~\bibnamefont{Schuster}},
  \bibinfo{author}{\bibfnamefont{W.}~\bibnamefont{Fontana}},
  \bibinfo{author}{\bibfnamefont{P.~F.} \bibnamefont{Stadler}},
  \bibnamefont{and} \bibinfo{author}{\bibfnamefont{I.~L.}
  \bibnamefont{Hofacker}}, \bibinfo{journal}{Proc.R. Soc. London B: Biological
  Sci.} \textbf{\bibinfo{volume}{255}}, \bibinfo{pages}{279}
  (\bibinfo{year}{1994}).

\bibitem[{\citenamefont{{Tinoco Jr.} and Bustamante}(1999)}]{Tinoco99JMB}
\bibinfo{author}{\bibfnamefont{I.}~\bibnamefont{{Tinoco Jr.}}}
  \bibnamefont{and}
  \bibinfo{author}{\bibfnamefont{C.}~\bibnamefont{Bustamante}},
  \bibinfo{journal}{J. Mol. Biol.} \textbf{\bibinfo{volume}{293}},
  \bibinfo{pages}{271} (\bibinfo{year}{1999}).

\bibitem[{\citenamefont{Thirumalai and Hyeon}(2005)}]{Thirum05Biochem}
\bibinfo{author}{\bibfnamefont{D.}~\bibnamefont{Thirumalai}} \bibnamefont{and}
  \bibinfo{author}{\bibfnamefont{C.}~\bibnamefont{Hyeon}},
  \bibinfo{journal}{Biochemistry} \textbf{\bibinfo{volume}{44}},
  \bibinfo{pages}{4957} (\bibinfo{year}{2005}).

\bibitem[{\citenamefont{Chen and Dill}(2000)}]{Chen00PNAS}
\bibinfo{author}{\bibfnamefont{S.~J.} \bibnamefont{Chen}} \bibnamefont{and}
  \bibinfo{author}{\bibfnamefont{K.~A.} \bibnamefont{Dill}},
  \bibinfo{journal}{Proc. Natl. Acad. Sci. U. S. A.}
  \textbf{\bibinfo{volume}{97}}, \bibinfo{pages}{646} (\bibinfo{year}{2000}).

\bibitem[{\citenamefont{Morcos et~al.}(2014)\citenamefont{Morcos, Schafer,
  Cheng, Onuchic, and Wolynes}}]{morcos2014PNAS}
\bibinfo{author}{\bibfnamefont{F.}~\bibnamefont{Morcos}},
  \bibinfo{author}{\bibfnamefont{N.~P.} \bibnamefont{Schafer}},
  \bibinfo{author}{\bibfnamefont{R.~R.} \bibnamefont{Cheng}},
  \bibinfo{author}{\bibfnamefont{J.~N.} \bibnamefont{Onuchic}},
  \bibnamefont{and} \bibinfo{author}{\bibfnamefont{P.~G.}
  \bibnamefont{Wolynes}}, \bibinfo{journal}{Proc. Natl. Acad. Sci. U. S. A.}
  \textbf{\bibinfo{volume}{111}}, \bibinfo{pages}{12408}
  (\bibinfo{year}{2014}).

\bibitem[{\citenamefont{Hyeon et~al.}(2006)\citenamefont{Hyeon, Dima, and
  Thirumalai}}]{Hyeon06JCP_2}
\bibinfo{author}{\bibfnamefont{C.}~\bibnamefont{Hyeon}},
  \bibinfo{author}{\bibfnamefont{R.~I.} \bibnamefont{Dima}}, \bibnamefont{and}
  \bibinfo{author}{\bibfnamefont{D.}~\bibnamefont{Thirumalai}},
  \bibinfo{journal}{J. Chem. Phys.} \textbf{\bibinfo{volume}{125}},
  \bibinfo{pages}{194905} (\bibinfo{year}{2006}).

\bibitem[{\citenamefont{Thirumalai et~al.}(2001)\citenamefont{Thirumalai, Lee,
  Woodson, and Klimov}}]{ThirumARPC01}
\bibinfo{author}{\bibfnamefont{D.}~\bibnamefont{Thirumalai}},
  \bibinfo{author}{\bibfnamefont{N.}~\bibnamefont{Lee}},
  \bibinfo{author}{\bibfnamefont{S.~A.} \bibnamefont{Woodson}},
  \bibnamefont{and} \bibinfo{author}{\bibfnamefont{D.~K.}
  \bibnamefont{Klimov}}, \bibinfo{journal}{Annu. Rev. Phys. Chem.}
  \textbf{\bibinfo{volume}{52}}, \bibinfo{pages}{751} (\bibinfo{year}{2001}).

\bibitem[{\citenamefont{Langer-Safer et~al.}(1982)\citenamefont{Langer-Safer,
  Levine, and Ward}}]{langer1982PNAS}
\bibinfo{author}{\bibfnamefont{P.~R.} \bibnamefont{Langer-Safer}},
  \bibinfo{author}{\bibfnamefont{M.}~\bibnamefont{Levine}}, \bibnamefont{and}
  \bibinfo{author}{\bibfnamefont{D.~C.} \bibnamefont{Ward}},
  \bibinfo{journal}{Proc. Natl. Acad. Sci. U. S. A.}
  \textbf{\bibinfo{volume}{79}}, \bibinfo{pages}{4381} (\bibinfo{year}{1982}).

\bibitem[{\citenamefont{Cremer and Cremer}(2001)}]{cremer2001NRG}
\bibinfo{author}{\bibfnamefont{T.}~\bibnamefont{Cremer}} \bibnamefont{and}
  \bibinfo{author}{\bibfnamefont{C.}~\bibnamefont{Cremer}},
  \bibinfo{journal}{Nature Rev. Genet.} \textbf{\bibinfo{volume}{2}},
  \bibinfo{pages}{292} (\bibinfo{year}{2001}).

\bibitem[{\citenamefont{Dekker et~al.}(2002)\citenamefont{Dekker, Rippe,
  Dekker, and Kleckner}}]{Dekker2002Science}
\bibinfo{author}{\bibfnamefont{J.}~\bibnamefont{Dekker}},
  \bibinfo{author}{\bibfnamefont{K.}~\bibnamefont{Rippe}},
  \bibinfo{author}{\bibfnamefont{M.}~\bibnamefont{Dekker}}, \bibnamefont{and}
  \bibinfo{author}{\bibfnamefont{N.}~\bibnamefont{Kleckner}},
  \bibinfo{journal}{Science} \textbf{\bibinfo{volume}{295}},
  \bibinfo{pages}{1306} (\bibinfo{year}{2002}).

\bibitem[{\citenamefont{Lieberman-Aiden
  et~al.}(2009)\citenamefont{Lieberman-Aiden, van Berkum, Williams, Imakaev,
  Ragoczy, Telling, Amit, Lajoie, Sabo, Dorschner et~al.}}]{lieberman09Science}
\bibinfo{author}{\bibfnamefont{E.}~\bibnamefont{Lieberman-Aiden}},
  \bibinfo{author}{\bibfnamefont{N.}~\bibnamefont{van Berkum}},
  \bibinfo{author}{\bibfnamefont{L.}~\bibnamefont{Williams}},
  \bibinfo{author}{\bibfnamefont{M.}~\bibnamefont{Imakaev}},
  \bibinfo{author}{\bibfnamefont{T.}~\bibnamefont{Ragoczy}},
  \bibinfo{author}{\bibfnamefont{A.}~\bibnamefont{Telling}},
  \bibinfo{author}{\bibfnamefont{I.}~\bibnamefont{Amit}},
  \bibinfo{author}{\bibfnamefont{B.}~\bibnamefont{Lajoie}},
  \bibinfo{author}{\bibfnamefont{P.}~\bibnamefont{Sabo}},
  \bibinfo{author}{\bibfnamefont{M.}~\bibnamefont{Dorschner}},
  \bibnamefont{et~al.}, \bibinfo{journal}{Science}
  \textbf{\bibinfo{volume}{326}}, \bibinfo{pages}{289} (\bibinfo{year}{2009}).

\bibitem[{\citenamefont{Dekker et~al.}(2013)\citenamefont{Dekker, Marti-Renom,
  and Mirny}}]{dekker2013NRG}
\bibinfo{author}{\bibfnamefont{J.}~\bibnamefont{Dekker}},
  \bibinfo{author}{\bibfnamefont{M.~A.} \bibnamefont{Marti-Renom}},
  \bibnamefont{and} \bibinfo{author}{\bibfnamefont{L.~A.} \bibnamefont{Mirny}},
  \bibinfo{journal}{Nat. Rev. Genetics} \textbf{\bibinfo{volume}{14}},
  \bibinfo{pages}{390} (\bibinfo{year}{2013}).

\bibitem[{\citenamefont{Zuin et~al.}(2014)\citenamefont{Zuin, Dixon, van~der
  Reijden, Ye, Kolovos, Brouwer, van~de Corput, van~de Werken, Knoch, van
  IJcken et~al.}}]{zuin2014PNAS}
\bibinfo{author}{\bibfnamefont{J.}~\bibnamefont{Zuin}},
  \bibinfo{author}{\bibfnamefont{J.~R.} \bibnamefont{Dixon}},
  \bibinfo{author}{\bibfnamefont{M.~I.} \bibnamefont{van~der Reijden}},
  \bibinfo{author}{\bibfnamefont{Z.}~\bibnamefont{Ye}},
  \bibinfo{author}{\bibfnamefont{P.}~\bibnamefont{Kolovos}},
  \bibinfo{author}{\bibfnamefont{R.~W.} \bibnamefont{Brouwer}},
  \bibinfo{author}{\bibfnamefont{M.~P.} \bibnamefont{van~de Corput}},
  \bibinfo{author}{\bibfnamefont{H.~J.} \bibnamefont{van~de Werken}},
  \bibinfo{author}{\bibfnamefont{T.~A.} \bibnamefont{Knoch}},
  \bibinfo{author}{\bibfnamefont{W.~F.} \bibnamefont{van IJcken}},
  \bibnamefont{et~al.}, \bibinfo{journal}{Proc. Natl. Acad. Sci. U. S. A.}
  \textbf{\bibinfo{volume}{111}}, \bibinfo{pages}{996} (\bibinfo{year}{2014}).

\bibitem[{\citenamefont{Grosberg and Khokhlov}(1994)}]{GrosbergBook}
\bibinfo{author}{\bibfnamefont{A.~Y.} \bibnamefont{Grosberg}} \bibnamefont{and}
  \bibinfo{author}{\bibfnamefont{A.~R.} \bibnamefont{Khokhlov}},
  \emph{\bibinfo{title}{Statistical Physics of Macromolecules}}
  (\bibinfo{publisher}{AIP Press}, \bibinfo{address}{New York},
  \bibinfo{year}{1994}).

\bibitem[{\citenamefont{Grosberg et~al.}(1988)\citenamefont{Grosberg, Nechaev,
  and Shakhnovich}}]{grosberg1988JP}
\bibinfo{author}{\bibfnamefont{A.}~\bibnamefont{Grosberg}},
  \bibinfo{author}{\bibfnamefont{S.}~\bibnamefont{Nechaev}}, \bibnamefont{and}
  \bibinfo{author}{\bibfnamefont{E.}~\bibnamefont{Shakhnovich}},
  \bibinfo{journal}{J. Phys.} \textbf{\bibinfo{volume}{49}},
  \bibinfo{pages}{2095} (\bibinfo{year}{1988}).

\bibitem[{\citenamefont{Mirny}(2011)}]{Mirny11ChromoRes}
\bibinfo{author}{\bibfnamefont{L.~A.} \bibnamefont{Mirny}},
  \bibinfo{journal}{Chromosome Res.} \textbf{\bibinfo{volume}{19}},
  \bibinfo{pages}{37} (\bibinfo{year}{2011}).

\bibitem[{\citenamefont{Bohn et~al.}(2007)\citenamefont{Bohn, Heermann, and van
  Driel}}]{2007Bohn051805}
\bibinfo{author}{\bibfnamefont{M.}~\bibnamefont{Bohn}},
  \bibinfo{author}{\bibfnamefont{D.~W.} \bibnamefont{Heermann}},
  \bibnamefont{and} \bibinfo{author}{\bibfnamefont{R.}~\bibnamefont{van
  Driel}}, \bibinfo{journal}{Phys. Rev. E} \textbf{\bibinfo{volume}{76}},
  \bibinfo{pages}{051805} (\bibinfo{year}{2007}).

\bibitem[{\citenamefont{Barbieri et~al.}(2012)\citenamefont{Barbieri, Chotalia,
  Fraser, Lavitas, Dostie, Pombo, and Nicodemi}}]{Barbieri12PNAS}
\bibinfo{author}{\bibfnamefont{M.}~\bibnamefont{Barbieri}},
  \bibinfo{author}{\bibfnamefont{M.}~\bibnamefont{Chotalia}},
  \bibinfo{author}{\bibfnamefont{J.}~\bibnamefont{Fraser}},
  \bibinfo{author}{\bibfnamefont{L.-M.} \bibnamefont{Lavitas}},
  \bibinfo{author}{\bibfnamefont{J.}~\bibnamefont{Dostie}},
  \bibinfo{author}{\bibfnamefont{A.}~\bibnamefont{Pombo}}, \bibnamefont{and}
  \bibinfo{author}{\bibfnamefont{M.}~\bibnamefont{Nicodemi}},
  \bibinfo{journal}{Proc. Natl. Acad. Sci. U. S. A.}
  \textbf{\bibinfo{volume}{109}}, \bibinfo{pages}{16173}
  (\bibinfo{year}{2012}).

\bibitem[{\citenamefont{Kang et~al.}(2015)\citenamefont{Kang, Yoon, Thirumalai,
  and Hyeon}}]{Kang2015PRL}
\bibinfo{author}{\bibfnamefont{H.}~\bibnamefont{Kang}},
  \bibinfo{author}{\bibfnamefont{Y.-G.} \bibnamefont{Yoon}},
  \bibinfo{author}{\bibfnamefont{D.}~\bibnamefont{Thirumalai}},
  \bibnamefont{and} \bibinfo{author}{\bibfnamefont{C.}~\bibnamefont{Hyeon}},
  \bibinfo{journal}{Phys. Rev. Lett.} \textbf{\bibinfo{volume}{115}},
  \bibinfo{pages}{198102} (\bibinfo{year}{2015}).

\bibitem[{\citenamefont{Pyetan et~al.}(2007)\citenamefont{Pyetan, Baram,
  Auerbach-Nevo, and Yonath}}]{Pyetan2007PAC}
\bibinfo{author}{\bibfnamefont{E.}~\bibnamefont{Pyetan}},
  \bibinfo{author}{\bibfnamefont{D.}~\bibnamefont{Baram}},
  \bibinfo{author}{\bibfnamefont{T.}~\bibnamefont{Auerbach-Nevo}},
  \bibnamefont{and} \bibinfo{author}{\bibfnamefont{A.}~\bibnamefont{Yonath}},
  \bibinfo{journal}{Pure Appl. Chem.} \textbf{\bibinfo{volume}{79}},
  \bibinfo{pages}{955} (\bibinfo{year}{2007}).

\bibitem[{\citenamefont{Guo et~al.}(2011)\citenamefont{Guo, Yuan, Xu, Feng,
  Liu, Chen, Sun, Yang, Lei, and Gao}}]{guo2011PNAS}
\bibinfo{author}{\bibfnamefont{Q.}~\bibnamefont{Guo}},
  \bibinfo{author}{\bibfnamefont{Y.}~\bibnamefont{Yuan}},
  \bibinfo{author}{\bibfnamefont{Y.}~\bibnamefont{Xu}},
  \bibinfo{author}{\bibfnamefont{B.}~\bibnamefont{Feng}},
  \bibinfo{author}{\bibfnamefont{L.}~\bibnamefont{Liu}},
  \bibinfo{author}{\bibfnamefont{K.}~\bibnamefont{Chen}},
  \bibinfo{author}{\bibfnamefont{M.}~\bibnamefont{Sun}},
  \bibinfo{author}{\bibfnamefont{Z.}~\bibnamefont{Yang}},
  \bibinfo{author}{\bibfnamefont{J.}~\bibnamefont{Lei}}, \bibnamefont{and}
  \bibinfo{author}{\bibfnamefont{N.}~\bibnamefont{Gao}},
  \bibinfo{journal}{Proc. Natl. Acad. Sci. U. S. A.}
  \textbf{\bibinfo{volume}{108}}, \bibinfo{pages}{13100}
  (\bibinfo{year}{2011}).

\bibitem[{\citenamefont{Golden et~al.}(2005)\citenamefont{Golden, Kim, and
  Chase}}]{Golden2005NSMB}
\bibinfo{author}{\bibfnamefont{B.~L.} \bibnamefont{Golden}},
  \bibinfo{author}{\bibfnamefont{H.}~\bibnamefont{Kim}}, \bibnamefont{and}
  \bibinfo{author}{\bibfnamefont{E.}~\bibnamefont{Chase}},
  \bibinfo{journal}{Nature. Struct. Mol. Biol.} \textbf{\bibinfo{volume}{12}},
  \bibinfo{pages}{82} (\bibinfo{year}{2005}).

\bibitem[{\citenamefont{Krasilnikov et~al.}(2004)\citenamefont{Krasilnikov,
  Xiao, Pan, and Mondrag{\'o}n}}]{krasilnikov2004Science}
\bibinfo{author}{\bibfnamefont{A.~S.} \bibnamefont{Krasilnikov}},
  \bibinfo{author}{\bibfnamefont{Y.}~\bibnamefont{Xiao}},
  \bibinfo{author}{\bibfnamefont{T.}~\bibnamefont{Pan}}, \bibnamefont{and}
  \bibinfo{author}{\bibfnamefont{A.}~\bibnamefont{Mondrag{\'o}n}},
  \bibinfo{journal}{Science} \textbf{\bibinfo{volume}{306}},
  \bibinfo{pages}{104} (\bibinfo{year}{2004}).

\bibitem[{\citenamefont{Thore et~al.}(2008)\citenamefont{Thore, Frick, and
  Ban}}]{Thore2008JACS}
\bibinfo{author}{\bibfnamefont{S.}~\bibnamefont{Thore}},
  \bibinfo{author}{\bibfnamefont{C.}~\bibnamefont{Frick}}, \bibnamefont{and}
  \bibinfo{author}{\bibfnamefont{N.}~\bibnamefont{Ban}}, \bibinfo{journal}{J.
  Am. Chem. Soc.} \textbf{\bibinfo{volume}{130}}, \bibinfo{pages}{8116}
  (\bibinfo{year}{2008}).

\bibitem[{\citenamefont{Ferguson et~al.}(2000)\citenamefont{Ferguson, Braun,
  Fiedler, Coulton, Diederichs, and Welte}}]{ferguson2000ProteinSci}
\bibinfo{author}{\bibfnamefont{A.~D.} \bibnamefont{Ferguson}},
  \bibinfo{author}{\bibfnamefont{V.}~\bibnamefont{Braun}},
  \bibinfo{author}{\bibfnamefont{H.-P.} \bibnamefont{Fiedler}},
  \bibinfo{author}{\bibfnamefont{J.~W.} \bibnamefont{Coulton}},
  \bibinfo{author}{\bibfnamefont{K.}~\bibnamefont{Diederichs}},
  \bibnamefont{and} \bibinfo{author}{\bibfnamefont{W.}~\bibnamefont{Welte}},
  \bibinfo{journal}{Protein Science} \textbf{\bibinfo{volume}{9}},
  \bibinfo{pages}{956} (\bibinfo{year}{2000}).

\bibitem[{\citenamefont{Otterbein et~al.}(2001)\citenamefont{Otterbein,
  Graceffa, and Dominguez}}]{2001Otterbein708}
\bibinfo{author}{\bibfnamefont{L.~R.} \bibnamefont{Otterbein}},
  \bibinfo{author}{\bibfnamefont{P.}~\bibnamefont{Graceffa}}, \bibnamefont{and}
  \bibinfo{author}{\bibfnamefont{R.}~\bibnamefont{Dominguez}},
  \bibinfo{journal}{Science} \textbf{\bibinfo{volume}{293}},
  \bibinfo{pages}{708} (\bibinfo{year}{2001}).

\bibitem[{\citenamefont{McLuskey et~al.}(2012)\citenamefont{McLuskey, Rudolf,
  Proto, Isaacs, Coombs, Moss, and Mottram}}]{2012McLuskey7469}
\bibinfo{author}{\bibfnamefont{K.}~\bibnamefont{McLuskey}},
  \bibinfo{author}{\bibfnamefont{J.}~\bibnamefont{Rudolf}},
  \bibinfo{author}{\bibfnamefont{W.~R.} \bibnamefont{Proto}},
  \bibinfo{author}{\bibfnamefont{N.~W.} \bibnamefont{Isaacs}},
  \bibinfo{author}{\bibfnamefont{G.~H.} \bibnamefont{Coombs}},
  \bibinfo{author}{\bibfnamefont{C.~X.} \bibnamefont{Moss}}, \bibnamefont{and}
  \bibinfo{author}{\bibfnamefont{J.~C.} \bibnamefont{Mottram}},
  \bibinfo{journal}{Proc. Natl. Acad. Sci. U. S. A.}
  \textbf{\bibinfo{volume}{109}}, \bibinfo{pages}{7469} (\bibinfo{year}{2012}).

\bibitem[{\citenamefont{Orm\"{o} et~al.}(1996)\citenamefont{Orm\"{o}, Cubitt,
  Kallio, Gross, Tsien, and Remington}}]{1996Ormo1392}
\bibinfo{author}{\bibfnamefont{M.}~\bibnamefont{Orm\"{o}}},
  \bibinfo{author}{\bibfnamefont{A.~B.} \bibnamefont{Cubitt}},
  \bibinfo{author}{\bibfnamefont{K.}~\bibnamefont{Kallio}},
  \bibinfo{author}{\bibfnamefont{L.~A.} \bibnamefont{Gross}},
  \bibinfo{author}{\bibfnamefont{R.~Y.} \bibnamefont{Tsien}}, \bibnamefont{and}
  \bibinfo{author}{\bibfnamefont{S.~J.} \bibnamefont{Remington}},
  \bibinfo{journal}{Science} \textbf{\bibinfo{volume}{273}},
  \bibinfo{pages}{1392} (\bibinfo{year}{1996}).

\bibitem[{\citenamefont{Weaver and Matthews}(1987)}]{1987Weaver189}
\bibinfo{author}{\bibfnamefont{L.}~\bibnamefont{Weaver}} \bibnamefont{and}
  \bibinfo{author}{\bibfnamefont{B.}~\bibnamefont{Matthews}},
  \bibinfo{journal}{J. Mol. Biol.} \textbf{\bibinfo{volume}{193}},
  \bibinfo{pages}{189 } (\bibinfo{year}{1987}).

\bibitem[{\citenamefont{Shank et~al.}(2010)\citenamefont{Shank, Cecconi, Dill,
  Marqusee, and Bustamante}}]{shank2010Nature}
\bibinfo{author}{\bibfnamefont{E.~A.} \bibnamefont{Shank}},
  \bibinfo{author}{\bibfnamefont{C.}~\bibnamefont{Cecconi}},
  \bibinfo{author}{\bibfnamefont{J.~W.} \bibnamefont{Dill}},
  \bibinfo{author}{\bibfnamefont{S.}~\bibnamefont{Marqusee}}, \bibnamefont{and}
  \bibinfo{author}{\bibfnamefont{C.}~\bibnamefont{Bustamante}},
  \bibinfo{journal}{Nature} \textbf{\bibinfo{volume}{465}},
  \bibinfo{pages}{637} (\bibinfo{year}{2010}).

\bibitem[{\citenamefont{Huang et~al.}(2003)\citenamefont{Huang, Lunin, Li,
  Suzuki, Sugiura, Miyazono, and Cygler}}]{Huang2003JMB}
\bibinfo{author}{\bibfnamefont{W.}~\bibnamefont{Huang}},
  \bibinfo{author}{\bibfnamefont{V.}~\bibnamefont{Lunin}},
  \bibinfo{author}{\bibfnamefont{Y.}~\bibnamefont{Li}},
  \bibinfo{author}{\bibfnamefont{S.}~\bibnamefont{Suzuki}},
  \bibinfo{author}{\bibfnamefont{N.}~\bibnamefont{Sugiura}},
  \bibinfo{author}{\bibfnamefont{H.}~\bibnamefont{Miyazono}}, \bibnamefont{and}
  \bibinfo{author}{\bibfnamefont{M.}~\bibnamefont{Cygler}},
  \bibinfo{journal}{J. Mol. Biol.} \textbf{\bibinfo{volume}{328}},
  \bibinfo{pages}{623} (\bibinfo{year}{2003}).

\bibitem[{\citenamefont{Batey et~al.}(1999)\citenamefont{Batey, Rambo, Doudna
  et~al.}}]{batey1999Angew}
\bibinfo{author}{\bibfnamefont{R.~T.} \bibnamefont{Batey}},
  \bibinfo{author}{\bibfnamefont{R.~P.} \bibnamefont{Rambo}},
  \bibinfo{author}{\bibfnamefont{J.~A.} \bibnamefont{Doudna}},
  \bibnamefont{et~al.}, \bibinfo{journal}{Angew. Chem. Int. Ed.}
  \textbf{\bibinfo{volume}{38}}, \bibinfo{pages}{2326} (\bibinfo{year}{1999}).

\bibitem[{\citenamefont{Nissen et~al.}(2001)\citenamefont{Nissen, Ippolito,
  Ban, Moore, and Steitz}}]{Nissen2001PNAS}
\bibinfo{author}{\bibfnamefont{P.}~\bibnamefont{Nissen}},
  \bibinfo{author}{\bibfnamefont{J.~A.} \bibnamefont{Ippolito}},
  \bibinfo{author}{\bibfnamefont{N.}~\bibnamefont{Ban}},
  \bibinfo{author}{\bibfnamefont{P.~B.} \bibnamefont{Moore}}, \bibnamefont{and}
  \bibinfo{author}{\bibfnamefont{T.~A.} \bibnamefont{Steitz}},
  \bibinfo{journal}{Proc. Natl. Acad. Sci. U. S. A.}
  \textbf{\bibinfo{volume}{98}}, \bibinfo{pages}{4899} (\bibinfo{year}{2001}).

\bibitem[{\citenamefont{Lua and Grosberg}(2006)}]{2006Luae45}
\bibinfo{author}{\bibfnamefont{R.~C.} \bibnamefont{Lua}} \bibnamefont{and}
  \bibinfo{author}{\bibfnamefont{A.~Y.} \bibnamefont{Grosberg}},
  \bibinfo{journal}{PLoS Comput Biol} \textbf{\bibinfo{volume}{2}},
  \bibinfo{pages}{e45} (\bibinfo{year}{2006}).

\bibitem[{\citenamefont{Halverson et~al.}(2014)\citenamefont{Halverson, Smrek,
  Kremer, and Grosberg}}]{Halverson14RPP}
\bibinfo{author}{\bibfnamefont{J.~D.} \bibnamefont{Halverson}},
  \bibinfo{author}{\bibfnamefont{J.}~\bibnamefont{Smrek}},
  \bibinfo{author}{\bibfnamefont{K.}~\bibnamefont{Kremer}}, \bibnamefont{and}
  \bibinfo{author}{\bibfnamefont{A.~Y.} \bibnamefont{Grosberg}},
  \bibinfo{journal}{Rep. Prog. Phys.} \textbf{\bibinfo{volume}{77}},
  \bibinfo{pages}{022601} (\bibinfo{year}{2014}).

\bibitem[{\citenamefont{Berezovsky et~al.}(2000)\citenamefont{Berezovsky,
  Grosberg, and Trifonov}}]{2000Berezovsky283}
\bibinfo{author}{\bibfnamefont{I.~N.} \bibnamefont{Berezovsky}},
  \bibinfo{author}{\bibfnamefont{A.~Y.} \bibnamefont{Grosberg}},
  \bibnamefont{and} \bibinfo{author}{\bibfnamefont{E.~N.}
  \bibnamefont{Trifonov}}, \bibinfo{journal}{FEBS Lett}
  \textbf{\bibinfo{volume}{466}}, \bibinfo{pages}{283} (\bibinfo{year}{2000}).

\bibitem[{\citenamefont{Sanborn et~al.}(2015)\citenamefont{Sanborn, Rao, Huang,
  Durand, Huntley, Jewett, Bochkov, Chinnappan, Cutkosky, Li
  et~al.}}]{sanborn2015PNAS}
\bibinfo{author}{\bibfnamefont{A.~L.} \bibnamefont{Sanborn}},
  \bibinfo{author}{\bibfnamefont{S.~S.} \bibnamefont{Rao}},
  \bibinfo{author}{\bibfnamefont{S.-C.} \bibnamefont{Huang}},
  \bibinfo{author}{\bibfnamefont{N.~C.} \bibnamefont{Durand}},
  \bibinfo{author}{\bibfnamefont{M.~H.} \bibnamefont{Huntley}},
  \bibinfo{author}{\bibfnamefont{A.~I.} \bibnamefont{Jewett}},
  \bibinfo{author}{\bibfnamefont{I.~D.} \bibnamefont{Bochkov}},
  \bibinfo{author}{\bibfnamefont{D.}~\bibnamefont{Chinnappan}},
  \bibinfo{author}{\bibfnamefont{A.}~\bibnamefont{Cutkosky}},
  \bibinfo{author}{\bibfnamefont{J.}~\bibnamefont{Li}}, \bibnamefont{et~al.},
  \bibinfo{journal}{Proc. Natl. Acad. Sci. U. S. A.}
  \textbf{\bibinfo{volume}{112}}, \bibinfo{pages}{E6456}
  (\bibinfo{year}{2015}).

\bibitem[{\citenamefont{Flory}(1969)}]{Florybook}
\bibinfo{author}{\bibfnamefont{P.~J.} \bibnamefont{Flory}},
  \emph{\bibinfo{title}{{Statistical Mechanics of Chain Molecules}}}, New York
  (\bibinfo{publisher}{Interscience Publishers}, \bibinfo{year}{1969}).

\bibitem[{\citenamefont{Flory}(1949)}]{Flory49JCP}
\bibinfo{author}{\bibfnamefont{P.~J.} \bibnamefont{Flory}},
  \bibinfo{journal}{J. Chem. Phys.} \textbf{\bibinfo{volume}{17}},
  \bibinfo{pages}{303} (\bibinfo{year}{1949}).

\bibitem[{\citenamefont{{de Gennes}}(1979)}]{deGennesbook}
\bibinfo{author}{\bibfnamefont{P.~G.} \bibnamefont{{de Gennes}}},
  \emph{\bibinfo{title}{{Scaling Concepts in Polymer Physics}}}
  (\bibinfo{publisher}{Cornell University Press}, \bibinfo{address}{Ithaca and
  London}, \bibinfo{year}{1979}).

\bibitem[{\citenamefont{Cacciuto and Luijten}(2006)}]{Cacciuto2006NanoLett}
\bibinfo{author}{\bibfnamefont{A.}~\bibnamefont{Cacciuto}} \bibnamefont{and}
  \bibinfo{author}{\bibfnamefont{E.}~\bibnamefont{Luijten}},
  \bibinfo{journal}{Nano Lett.} \textbf{\bibinfo{volume}{6}},
  \bibinfo{pages}{901} (\bibinfo{year}{2006}).

\bibitem[{\citenamefont{Onoa et~al.}(2003)\citenamefont{Onoa, Dumont, Liphardt,
  Smith, Tinoco, and Bustamante}}]{Onoa2003Science}
\bibinfo{author}{\bibfnamefont{B.}~\bibnamefont{Onoa}},
  \bibinfo{author}{\bibfnamefont{S.}~\bibnamefont{Dumont}},
  \bibinfo{author}{\bibfnamefont{J.}~\bibnamefont{Liphardt}},
  \bibinfo{author}{\bibfnamefont{S.~B.} \bibnamefont{Smith}},
  \bibinfo{author}{\bibfnamefont{I.}~\bibnamefont{Tinoco}}, \bibnamefont{and}
  \bibinfo{author}{\bibfnamefont{C.}~\bibnamefont{Bustamante}},
  \bibinfo{journal}{Science} \textbf{\bibinfo{volume}{299}},
  \bibinfo{pages}{1892} (\bibinfo{year}{2003}).

\bibitem[{\citenamefont{Mickler et~al.}(2007)\citenamefont{Mickler, Dima,
  Dietz, Hyeon, Thirumalai, and Rief}}]{Mickler07PNAS}
\bibinfo{author}{\bibfnamefont{M.}~\bibnamefont{Mickler}},
  \bibinfo{author}{\bibfnamefont{R.~I.} \bibnamefont{Dima}},
  \bibinfo{author}{\bibfnamefont{H.}~\bibnamefont{Dietz}},
  \bibinfo{author}{\bibfnamefont{C.}~\bibnamefont{Hyeon}},
  \bibinfo{author}{\bibfnamefont{D.}~\bibnamefont{Thirumalai}},
  \bibnamefont{and} \bibinfo{author}{\bibfnamefont{M.}~\bibnamefont{Rief}},
  \bibinfo{journal}{Proc. Natl. Acad. Sci. U. S. A.}
  \textbf{\bibinfo{volume}{104}}, \bibinfo{pages}{20268}
  (\bibinfo{year}{2007}).

\bibitem[{\citenamefont{Micheletti et~al.}(2015)\citenamefont{Micheletti,
  Di~Stefano, and Orland}}]{2015Micheletti2052}
\bibinfo{author}{\bibfnamefont{C.}~\bibnamefont{Micheletti}},
  \bibinfo{author}{\bibfnamefont{M.}~\bibnamefont{Di~Stefano}},
  \bibnamefont{and} \bibinfo{author}{\bibfnamefont{H.}~\bibnamefont{Orland}},
  \bibinfo{journal}{Proc. Natl. Acad. Sci. U. S. A.}
  \textbf{\bibinfo{volume}{112}}, \bibinfo{pages}{2052} (\bibinfo{year}{2015}).

\bibitem[{\citenamefont{Burton et~al.}(2015)\citenamefont{Burton, Di~Stefano,
  Lehman, Orland, and Micheletti}}]{burton2015RNAbiol}
\bibinfo{author}{\bibfnamefont{A.~S.} \bibnamefont{Burton}},
  \bibinfo{author}{\bibfnamefont{M.}~\bibnamefont{Di~Stefano}},
  \bibinfo{author}{\bibfnamefont{N.}~\bibnamefont{Lehman}},
  \bibinfo{author}{\bibfnamefont{H.}~\bibnamefont{Orland}}, \bibnamefont{and}
  \bibinfo{author}{\bibfnamefont{C.}~\bibnamefont{Micheletti}},
  \bibinfo{journal}{RNA Biology} \textbf{\bibinfo{volume}{13}},
  \bibinfo{pages}{134} (\bibinfo{year}{2015}).

\bibitem[{\citenamefont{Noel et~al.}(2010)\citenamefont{Noel, Su{\l}kowska, and
  Onuchic}}]{noel2010PNAS}
\bibinfo{author}{\bibfnamefont{J.~K.} \bibnamefont{Noel}},
  \bibinfo{author}{\bibfnamefont{J.~I.} \bibnamefont{Su{\l}kowska}},
  \bibnamefont{and} \bibinfo{author}{\bibfnamefont{J.~N.}
  \bibnamefont{Onuchic}}, \bibinfo{journal}{Proc. Natl. Acad. Sci. U. S. A.}
  \textbf{\bibinfo{volume}{107}}, \bibinfo{pages}{15403}
  (\bibinfo{year}{2010}).

\bibitem[{\citenamefont{Grosberg}(2000)}]{grosberg2000PRL}
\bibinfo{author}{\bibfnamefont{A.~Y.} \bibnamefont{Grosberg}},
  \bibinfo{journal}{Phys. Rev. Lett.} \textbf{\bibinfo{volume}{85}},
  \bibinfo{pages}{3858} (\bibinfo{year}{2000}).

\bibitem[{\citenamefont{Imakaev et~al.}(2015)\citenamefont{Imakaev, Tchourine,
  Nechaev, and Mirny}}]{imakaev2015SoftMatter}
\bibinfo{author}{\bibfnamefont{M.~V.} \bibnamefont{Imakaev}},
  \bibinfo{author}{\bibfnamefont{K.~M.} \bibnamefont{Tchourine}},
  \bibinfo{author}{\bibfnamefont{S.~K.} \bibnamefont{Nechaev}},
  \bibnamefont{and} \bibinfo{author}{\bibfnamefont{L.~A.} \bibnamefont{Mirny}},
  \bibinfo{journal}{Soft matter} \textbf{\bibinfo{volume}{11}},
  \bibinfo{pages}{665} (\bibinfo{year}{2015}).

\bibitem[{\citenamefont{Thirumalai and Hyeon}(2008)}]{Thirumalai08Noncoding}
\bibinfo{author}{\bibfnamefont{D.}~\bibnamefont{Thirumalai}} \bibnamefont{and}
  \bibinfo{author}{\bibfnamefont{C.}~\bibnamefont{Hyeon}},
  \emph{\bibinfo{title}{Non-Protein Coding RNAs}}
  (\bibinfo{publisher}{Springer}, \bibinfo{year}{2008}), chap.
  \bibinfo{chapter}{{Theory of RNA Folding: From Hairpins to Ribozymes}}.

\bibitem[{\citenamefont{Thirumalai}(1995)}]{Thirum95JPI}
\bibinfo{author}{\bibfnamefont{D.}~\bibnamefont{Thirumalai}},
  \bibinfo{journal}{J. Phys. I (Fr.)} \textbf{\bibinfo{volume}{5}},
  \bibinfo{pages}{1457} (\bibinfo{year}{1995}).

\bibitem[{\citenamefont{Greenleaf et~al.}(2008)\citenamefont{Greenleaf, Frieda,
  Foster, Woodside, and Block}}]{Greenleaf08Science}
\bibinfo{author}{\bibfnamefont{W.~J.} \bibnamefont{Greenleaf}},
  \bibinfo{author}{\bibfnamefont{K.~L.} \bibnamefont{Frieda}},
  \bibinfo{author}{\bibfnamefont{D.~A.~N.} \bibnamefont{Foster}},
  \bibinfo{author}{\bibfnamefont{M.~T.} \bibnamefont{Woodside}},
  \bibnamefont{and} \bibinfo{author}{\bibfnamefont{S.~M.} \bibnamefont{Block}},
  \bibinfo{journal}{Science} \textbf{\bibinfo{volume}{319}},
  \bibinfo{pages}{630} (\bibinfo{year}{2008}).

\bibitem[{\citenamefont{Treiber and Williamson}(2001)}]{Treiber01COSB}
\bibinfo{author}{\bibfnamefont{D.~K.} \bibnamefont{Treiber}} \bibnamefont{and}
  \bibinfo{author}{\bibfnamefont{J.~R.} \bibnamefont{Williamson}},
  \bibinfo{journal}{Curr. Opin. Struct. Biol.} \textbf{\bibinfo{volume}{11}},
  \bibinfo{pages}{309} (\bibinfo{year}{2001}).

\bibitem[{\citenamefont{Repsilber et~al.}(1999)\citenamefont{Repsilber, Wiese,
  Rachen, Schroeder, Riesner, and Steger}}]{Repsilber1999RNA}
\bibinfo{author}{\bibfnamefont{D.}~\bibnamefont{Repsilber}},
  \bibinfo{author}{\bibfnamefont{S.}~\bibnamefont{Wiese}},
  \bibinfo{author}{\bibfnamefont{M.}~\bibnamefont{Rachen}},
  \bibinfo{author}{\bibfnamefont{A.~W.} \bibnamefont{Schroeder}},
  \bibinfo{author}{\bibfnamefont{D.}~\bibnamefont{Riesner}}, \bibnamefont{and}
  \bibinfo{author}{\bibfnamefont{G.}~\bibnamefont{Steger}},
  \bibinfo{journal}{RNA} \textbf{\bibinfo{volume}{5}}, \bibinfo{pages}{574}
  (\bibinfo{year}{1999}).

\bibitem[{\citenamefont{Lutz et~al.}(2013)\citenamefont{Lutz, Faber, Verma,
  Klumpp, and Schug}}]{Lutz2013NAR}
\bibinfo{author}{\bibfnamefont{B.}~\bibnamefont{Lutz}},
  \bibinfo{author}{\bibfnamefont{M.}~\bibnamefont{Faber}},
  \bibinfo{author}{\bibfnamefont{A.}~\bibnamefont{Verma}},
  \bibinfo{author}{\bibfnamefont{S.}~\bibnamefont{Klumpp}}, \bibnamefont{and}
  \bibinfo{author}{\bibfnamefont{A.}~\bibnamefont{Schug}},
  \bibinfo{journal}{Nucleic Acids Res.} \textbf{\bibinfo{volume}{42}},
  \bibinfo{pages}{2687} (\bibinfo{year}{2013}).

\bibitem[{\citenamefont{Wu and {Tinoco, Jr}}(1998)}]{Wu98PNAS}
\bibinfo{author}{\bibfnamefont{M.}~\bibnamefont{Wu}} \bibnamefont{and}
  \bibinfo{author}{\bibfnamefont{I.}~\bibnamefont{{Tinoco, Jr}}},
  \bibinfo{journal}{Proc. Natl. Acad. Sci. U. S. A.}
  \textbf{\bibinfo{volume}{95}}, \bibinfo{pages}{11555} (\bibinfo{year}{1998}).

\bibitem[{\citenamefont{Koculi et~al.}(2012)\citenamefont{Koculi, Cho, Desai,
  Thirumalai, and Woodson}}]{Koculi2012NAR}
\bibinfo{author}{\bibfnamefont{E.}~\bibnamefont{Koculi}},
  \bibinfo{author}{\bibfnamefont{S.~S.} \bibnamefont{Cho}},
  \bibinfo{author}{\bibfnamefont{R.}~\bibnamefont{Desai}},
  \bibinfo{author}{\bibfnamefont{D.}~\bibnamefont{Thirumalai}},
  \bibnamefont{and} \bibinfo{author}{\bibfnamefont{S.~A.}
  \bibnamefont{Woodson}}, \bibinfo{journal}{Nucleic Acids Res.}
  \textbf{\bibinfo{volume}{40}}, \bibinfo{pages}{8011} (\bibinfo{year}{2012}).

\bibitem[{\citenamefont{Montange and Batey}(2008)}]{Montange08ARB}
\bibinfo{author}{\bibfnamefont{R.~K.} \bibnamefont{Montange}} \bibnamefont{and}
  \bibinfo{author}{\bibfnamefont{R.}~\bibnamefont{Batey}},
  \bibinfo{journal}{Annu. Rev Biophys.} \textbf{\bibinfo{volume}{37}},
  \bibinfo{pages}{117} (\bibinfo{year}{2008}).

\bibitem[{\citenamefont{Russell et~al.}(2013)\citenamefont{Russell,
  Jarmoskaite, and Lambowitz}}]{Russell2013RNAbiology}
\bibinfo{author}{\bibfnamefont{R.}~\bibnamefont{Russell}},
  \bibinfo{author}{\bibfnamefont{I.}~\bibnamefont{Jarmoskaite}},
  \bibnamefont{and} \bibinfo{author}{\bibfnamefont{A.~M.}
  \bibnamefont{Lambowitz}}, \bibinfo{journal}{RNA Biology}
  \textbf{\bibinfo{volume}{10}}, \bibinfo{pages}{44} (\bibinfo{year}{2013}).

\bibitem[{\citenamefont{Al-Hashimi and Walter}(2008)}]{AlHashimi08COSB}
\bibinfo{author}{\bibfnamefont{H.}~\bibnamefont{Al-Hashimi}} \bibnamefont{and}
  \bibinfo{author}{\bibfnamefont{N.}~\bibnamefont{Walter}},
  \bibinfo{journal}{Curr. Opin. Struct. Biol.} \textbf{\bibinfo{volume}{18}},
  \bibinfo{pages}{321} (\bibinfo{year}{2008}).

\bibitem[{\citenamefont{Solomatin et~al.}(2010)\citenamefont{Solomatin,
  Greenfeld, Chu, and Herschlag}}]{Solomatin10Nature}
\bibinfo{author}{\bibfnamefont{S.~V.} \bibnamefont{Solomatin}},
  \bibinfo{author}{\bibfnamefont{M.}~\bibnamefont{Greenfeld}},
  \bibinfo{author}{\bibfnamefont{S.}~\bibnamefont{Chu}}, \bibnamefont{and}
  \bibinfo{author}{\bibfnamefont{D.}~\bibnamefont{Herschlag}},
  \bibinfo{journal}{Nature} \textbf{\bibinfo{volume}{463}},
  \bibinfo{pages}{681} (\bibinfo{year}{2010}).

\bibitem[{\citenamefont{Hyeon et~al.}(2012)\citenamefont{Hyeon, Lee, Yoon,
  Hohng, and Thirumalai}}]{Hyeon2012NatureChem}
\bibinfo{author}{\bibfnamefont{C.}~\bibnamefont{Hyeon}},
  \bibinfo{author}{\bibfnamefont{J.}~\bibnamefont{Lee}},
  \bibinfo{author}{\bibfnamefont{J.}~\bibnamefont{Yoon}},
  \bibinfo{author}{\bibfnamefont{S.}~\bibnamefont{Hohng}}, \bibnamefont{and}
  \bibinfo{author}{\bibfnamefont{D.}~\bibnamefont{Thirumalai}},
  \bibinfo{journal}{Nat. Chem.} \textbf{\bibinfo{volume}{4}},
  \bibinfo{pages}{907} (\bibinfo{year}{2012}).

\bibitem[{\citenamefont{Hyeon et~al.}(2014)\citenamefont{Hyeon, Hinczewski, and
  Thirumalai}}]{Hyeon14PRL}
\bibinfo{author}{\bibfnamefont{C.}~\bibnamefont{Hyeon}},
  \bibinfo{author}{\bibfnamefont{M.}~\bibnamefont{Hinczewski}},
  \bibnamefont{and}
  \bibinfo{author}{\bibfnamefont{D.}~\bibnamefont{Thirumalai}},
  \bibinfo{journal}{Phys. Rev. Lett.} \textbf{\bibinfo{volume}{112}},
  \bibinfo{pages}{138101} (\bibinfo{year}{2014}).

\bibitem[{\citenamefont{Rivas and Eddy}(1999)}]{Rivas99JMB}
\bibinfo{author}{\bibfnamefont{E.}~\bibnamefont{Rivas}} \bibnamefont{and}
  \bibinfo{author}{\bibfnamefont{S.}~\bibnamefont{Eddy}}, \bibinfo{journal}{J.
  Mol. Biol.} \textbf{\bibinfo{volume}{285}}, \bibinfo{pages}{2053}
  (\bibinfo{year}{1999}), ISSN \bibinfo{issn}{0022-2836}.

\bibitem[{\citenamefont{Hofacker}(2003)}]{hofacker03NAR}
\bibinfo{author}{\bibfnamefont{I.}~\bibnamefont{Hofacker}},
  \bibinfo{journal}{Nucleic Acids Res.} \textbf{\bibinfo{volume}{31}},
  \bibinfo{pages}{3429} (\bibinfo{year}{2003}), ISSN \bibinfo{issn}{0305-1048}.

\bibitem[{\citenamefont{Zuker}(2003)}]{Zuker03NAR}
\bibinfo{author}{\bibfnamefont{M.}~\bibnamefont{Zuker}},
  \bibinfo{journal}{Nucleic Acids Res.} \textbf{\bibinfo{volume}{31}},
  \bibinfo{pages}{3406} (\bibinfo{year}{2003}).

\bibitem[{\citenamefont{Gutell et~al.}(2002)\citenamefont{Gutell, Lee, and
  Cannone}}]{GutellCOSB02}
\bibinfo{author}{\bibfnamefont{R.~R.} \bibnamefont{Gutell}},
  \bibinfo{author}{\bibfnamefont{J.~C.} \bibnamefont{Lee}}, \bibnamefont{and}
  \bibinfo{author}{\bibfnamefont{J.~J.} \bibnamefont{Cannone}},
  \bibinfo{journal}{Curr. Opin. Struct. Biol.} \textbf{\bibinfo{volume}{12}},
  \bibinfo{pages}{301} (\bibinfo{year}{2002}).

\bibitem[{\citenamefont{Mathews et~al.}(1999)\citenamefont{Mathews, Sabina,
  Zuker, and Turner}}]{MathewsJMB99}
\bibinfo{author}{\bibfnamefont{D.}~\bibnamefont{Mathews}},
  \bibinfo{author}{\bibfnamefont{J.}~\bibnamefont{Sabina}},
  \bibinfo{author}{\bibfnamefont{M.}~\bibnamefont{Zuker}}, \bibnamefont{and}
  \bibinfo{author}{\bibfnamefont{D.}~\bibnamefont{Turner}},
  \bibinfo{journal}{J. Mol. Biol.} \textbf{\bibinfo{volume}{288}},
  \bibinfo{pages}{911} (\bibinfo{year}{1999}).

\bibitem[{\citenamefont{Popot et~al.}(1987)\citenamefont{Popot, Gerchman, and
  Engelman}}]{Popot1987JMB}
\bibinfo{author}{\bibfnamefont{J.-L.} \bibnamefont{Popot}},
  \bibinfo{author}{\bibfnamefont{S.-E.} \bibnamefont{Gerchman}},
  \bibnamefont{and} \bibinfo{author}{\bibfnamefont{D.~M.}
  \bibnamefont{Engelman}}, \bibinfo{journal}{J. Mol. Biol.}
  \textbf{\bibinfo{volume}{198}}, \bibinfo{pages}{655} (\bibinfo{year}{1987}).

\bibitem[{\citenamefont{Bowie}(2005)}]{Bowie05Nature}
\bibinfo{author}{\bibfnamefont{J.~U.} \bibnamefont{Bowie}},
  \bibinfo{journal}{Nature} \textbf{\bibinfo{volume}{438}},
  \bibinfo{pages}{581} (\bibinfo{year}{2005}).

\bibitem[{\citenamefont{Kedrov et~al.}(2004)\citenamefont{Kedrov, Ziegler,
  Janovjak, K{\"u}hlbrandt, and M{\"u}ller}}]{kedrov2004JMB}
\bibinfo{author}{\bibfnamefont{A.}~\bibnamefont{Kedrov}},
  \bibinfo{author}{\bibfnamefont{C.}~\bibnamefont{Ziegler}},
  \bibinfo{author}{\bibfnamefont{H.}~\bibnamefont{Janovjak}},
  \bibinfo{author}{\bibfnamefont{W.}~\bibnamefont{K{\"u}hlbrandt}},
  \bibnamefont{and} \bibinfo{author}{\bibfnamefont{D.~J.}
  \bibnamefont{M{\"u}ller}}, \bibinfo{journal}{J. Mol. Biol.}
  \textbf{\bibinfo{volume}{340}}, \bibinfo{pages}{1143} (\bibinfo{year}{2004}).

\bibitem[{\citenamefont{Min et~al.}(2015)\citenamefont{Min, Jefferson, Bowie,
  and Yoon}}]{Min15NCB}
\bibinfo{author}{\bibfnamefont{D.}~\bibnamefont{Min}},
  \bibinfo{author}{\bibfnamefont{R.}~\bibnamefont{Jefferson}},
  \bibinfo{author}{\bibfnamefont{J.}~\bibnamefont{Bowie}}, \bibnamefont{and}
  \bibinfo{author}{\bibfnamefont{T.-Y.} \bibnamefont{Yoon}},
  \bibinfo{journal}{Nat. Chem. Biol.} \textbf{\bibinfo{volume}{11}},
  \bibinfo{pages}{981} (\bibinfo{year}{2015}).

\bibitem[{\citenamefont{Palmer}(1982)}]{Palmer82AP}
\bibinfo{author}{\bibfnamefont{R.}~\bibnamefont{Palmer}},
  \bibinfo{journal}{Adv. Phys.} \textbf{\bibinfo{volume}{31}},
  \bibinfo{pages}{669} (\bibinfo{year}{1982}).

\bibitem[{\citenamefont{Rinn and Chang}(2012)}]{rinn2012ARBiochem}
\bibinfo{author}{\bibfnamefont{J.~L.} \bibnamefont{Rinn}} \bibnamefont{and}
  \bibinfo{author}{\bibfnamefont{H.~Y.} \bibnamefont{Chang}},
  \bibinfo{journal}{Annu. Rev. Biochem.} \textbf{\bibinfo{volume}{81}}
  (\bibinfo{year}{2012}).

\bibitem[{\citenamefont{Carninci et~al.}(2005)\citenamefont{Carninci, Kasukawa,
  Katayama, Gough, Frith, Maeda, Oyama, Ravasi, Lenhard, Wells
  et~al.}}]{carninci2005Science}
\bibinfo{author}{\bibfnamefont{P.}~\bibnamefont{Carninci}},
  \bibinfo{author}{\bibfnamefont{T.}~\bibnamefont{Kasukawa}},
  \bibinfo{author}{\bibfnamefont{S.}~\bibnamefont{Katayama}},
  \bibinfo{author}{\bibfnamefont{J.}~\bibnamefont{Gough}},
  \bibinfo{author}{\bibfnamefont{M.}~\bibnamefont{Frith}},
  \bibinfo{author}{\bibfnamefont{N.}~\bibnamefont{Maeda}},
  \bibinfo{author}{\bibfnamefont{R.}~\bibnamefont{Oyama}},
  \bibinfo{author}{\bibfnamefont{T.}~\bibnamefont{Ravasi}},
  \bibinfo{author}{\bibfnamefont{B.}~\bibnamefont{Lenhard}},
  \bibinfo{author}{\bibfnamefont{C.}~\bibnamefont{Wells}},
  \bibnamefont{et~al.}, \bibinfo{journal}{Science}
  \textbf{\bibinfo{volume}{309}}, \bibinfo{pages}{1559} (\bibinfo{year}{2005}).

\bibitem[{\citenamefont{Friedman and O'Shaughnessy}(1991)}]{FriedmanJPII91}
\bibinfo{author}{\bibfnamefont{B.}~\bibnamefont{Friedman}} \bibnamefont{and}
  \bibinfo{author}{\bibfnamefont{B.}~\bibnamefont{O'Shaughnessy}},
  \bibinfo{journal}{J. Phys. II} \textbf{\bibinfo{volume}{1}},
  \bibinfo{pages}{471} (\bibinfo{year}{1991}).

\end{thebibliography}

\end{document}